\documentclass[journal]{IEEEtran}
\usepackage[dvipsnames]{xcolor}
\usepackage{amsmath,amsfonts}
\usepackage{algorithmic}
\usepackage{array}
\usepackage[caption=false,font=normalsize,labelfont=sf,textfont=sf]{subfig}
\usepackage{textcomp}
\usepackage{url}
\usepackage{verbatim}
\usepackage{graphicx}

\usepackage{float}
\usepackage{algorithmic}
\usepackage[moderate, tracking=normal, leading=normal]{savetrees}
\usepackage{graphicx}
\usepackage{textcomp}
\usepackage{colortbl}
\usepackage{booktabs}
\usepackage{tikz}
\usepackage{fontawesome5}
\usetikzlibrary{shapes.geometric}
\usetikzlibrary{shapes.arrows}
\usepackage{array}
\usepackage{pifont}
\usepackage{multirow}
\usepackage{pgfplots}
\usepackage{adjustbox}
\usepackage{placeins}
\usepackage{dblfloatfix}
\usepackage{caption}

\usetikzlibrary{patterns}

\usepackage{soul}

\usepackage{glossaries}
\setacronymstyle{long-short}
\newacronym{3gpp}{3GPP}{3rd Generation Partnership Project}
\newacronym{4g}{4G}{4th generation}
\newacronym{5g}{5G}{5th generation}
\newacronym{6g}{6G}{6th generation}
\newacronym{5gc}{5GC}{5G Core}
\newacronym{adc}{ADC}{Analog to Digital Converter}
\newacronym{aerpaw}{AERPAW}{Aerial Experimentation and Research Platform for Advanced Wireless}
\newacronym{ai}{AI}{Artificial Intelligence}
\newacronym{aimd}{AIMD}{Additive Increase Multiplicative Decrease}
\newacronym{am}{AM}{Acknowledged Mode}
\newacronym{amc}{AMC}{Adaptive Modulation and Coding}
\newacronym{amf}{AMF}{Access and Mobility Management Function}
\newacronym{aops}{AOPS}{Adaptive Order Prediction Scheduling}
\newacronym{api}{API}{Application Programming Interface}
\newacronym{apn}{APN}{Access Point Name}
\newacronym{ap}{AP}{Application Protocol}
\newacronym{aqm}{AQM}{Active Queue Management}
\newacronym{ausf}{AUSF}{Authentication Server Function}
\newacronym{avc}{AVC}{Advanced Video Coding}
\newacronym{awgn}{AGWN}{Additive White Gaussian Noise}
\newacronym{balia}{BALIA}{Balanced Link Adaptation Algorithm}
\newacronym{bbu}{BBU}{Base Band Unit}
\newacronym{bdp}{BDP}{Bandwidth-Delay Product}
\newacronym{ber}{BER}{Bit Error Rate}
\newacronym{bf}{BF}{Beamforming}
\newacronym{bler}{BLER}{Block Error Rate}
\newacronym{brr}{BRR}{Bayesian Ridge Regressor}
\newacronym{bs}{BS}{Base Station}
\newacronym{bsr}{BSR}{Buffer Status Report}
\newacronym{bss}{BSS}{Business Support System}
\newacronym{ca}{CA}{Carrier Aggregation}
\newacronym{caas}{CaaS}{Connectivity-as-a-Service}
\newacronym{cb}{CB}{Code Block}
\newacronym{cc}{CC}{Congestion Control}
\newacronym{ccid}{CCID}{Congestion Control ID}
\newacronym{cco}{CC}{Carrier Component}
\newacronym{cdd}{CDD}{Cyclic Delay Diversity}
\newacronym{cdf}{CDF}{Cumulative Distribution Function}
\newacronym{cdn}{CDN}{Content Distribution Network}
\newacronym{cli}{CLI}{Command-line Interface}
\newacronym{cn}{CN}{Core Network}
\newacronym{codel}{CoDel}{Controlled Delay Management}
\newacronym{comac}{COMAC}{Converged Multi-Access and Core}
\newacronym{cord}{CORD}{Central Office Re-architected as a Datacenter}
\newacronym{cornet}{CORNET}{COgnitive Radio NETwork}
\newacronym{cosmos}{COSMOS}{Cloud Enhanced Open Software Defined Mobile Wireless Testbed for City-Scale Deployment}
\newacronym{cots}{COTS}{Commercial Off-the-Shelf}
\newacronym{cp}{CP}{Control Plane}
\newacronym{cyp}{CP}{Cyclic Prefix}
\newacronym{up}{UP}{User Plane}
\newacronym{cpu}{CPU}{Central Processing Unit}
\newacronym{cqi}{CQI}{Channel Quality Information}
\newacronym{cr}{CR}{Cognitive Radio}
\newacronym{cran}{CRAN}{Cloud \gls{ran}}
\newacronym{crs}{CRS}{Cell Reference Signal}
\newacronym{csi}{CSI}{Channel State Information}
\newacronym{csirs}{CSI-RS}{Channel State Information - Reference Signal}
\newacronym{cu}{CU}{Central Unit}
\newacronym{d2tcp}{D$^2$TCP}{Deadline-aware Data center TCP}
\newacronym{d3}{D$^3$}{Deadline-Driven Delivery}
\newacronym{dac}{DAC}{Digital to Analog Converter}
\newacronym{dag}{DAG}{Directed Acyclic Graph}
\newacronym{das}{DAS}{Distributed Antenna System}
\newacronym{dash}{DASH}{Dynamic Adaptive Streaming over HTTP}
\newacronym{dc}{DC}{Dual Connectivity}
\newacronym{dccp}{DCCP}{Datagram Congestion Control Protocol}
\newacronym{dce}{DCE}{Direct Code Execution}
\newacronym{dci}{DCI}{Downlink Control Information}
\newacronym{dctcp}{DCTCP}{Data Center TCP}
\newacronym{dl}{DL}{Downlink}
\newacronym{dmr}{DMR}{Deadline Miss Ratio}
\newacronym{dmrs}{DMRS}{DeModulation Reference Signal}
\newacronym{drlcc}{DRL-CC}{Deep Reinforcement Learning Congestion Control}
\newacronym{drs}{DRS}{Discovery Reference Signal}
\newacronym{du}{DU}{Distributed Unit}
\newacronym{e2e}{E2E}{end-to-end}
\newacronym{earfcn}{EARFCN}{E-UTRA Absolute Radio Frequency Channel Number}
\newacronym{ecaas}{ECaaS}{Edge-Cloud-as-a-Service}
\newacronym{ecn}{ECN}{Explicit Congestion Notification}
\newacronym{edf}{EDF}{Earliest Deadline First}
\newacronym{embb}{eMBB}{Enhanced Mobile Broadband}
\newacronym{empower}{EMPOWER}{EMpowering transatlantic PlatfOrms for advanced WirEless Research}
\newacronym{enb}{eNB}{evolved Node Base}
\newacronym{endc}{EN-DC}{E-UTRAN-\gls{nr} \gls{dc}}
\newacronym{epc}{EPC}{Evolved Packet Core}
\newacronym{eps}{EPS}{Evolved Packet System}
\newacronym{es}{ES}{Edge Server}
\newacronym{etsi}{ETSI}{European Telecommunications Standards Institute}
\newacronym[firstplural=Estimated Times of Arrival (ETAs)]{eta}{ETA}{Estimated Time of Arrival}
\newacronym{eutran}{E-UTRAN}{Evolved Universal Terrestrial Access Network}
\newacronym{faas}{FaaS}{Function-as-a-Service}
\newacronym{fapi}{FAPI}{Functional Application Platform Interface}
\newacronym{fdd}{FDD}{Frequency Division Duplexing}
\newacronym{fdm}{FDM}{Frequency Division Multiplexing}
\newacronym{fdma}{FDMA}{Frequency Division Multiple Access}
\newacronym{fed4fire}{FED4FIRE+}{Federation 4 Future Internet Research and Experimentation Plus}
\newacronym{fir}{FIR}{Finite Impulse Response}
\newacronym{fit}{FIT}{Future \acrlong{iot}}
\newacronym{fpga}{FPGA}{Field Programmable Gate Array}
\newacronym{fr2}{FR2}{Frequency Range 2}
\newacronym{fs}{FS}{Fast Switching}
\newacronym{fscc}{FSCC}{Flow Sharing Congestion Control}
\newacronym{ftp}{FTP}{File Transfer Protocol}
\newacronym{fw}{FW}{Flow Window}
\newacronym{ge}{GE}{Gaussian Elimination}
\newacronym{gnb}{gNB}{Next Generation Node Base}
\newacronym{gop}{GOP}{Group of Pictures}
\newacronym{gpr}{GPR}{Gaussian Process Regressor}
\newacronym{gpu}{GPU}{Graphics Processing Unit}
\newacronym{gtp}{GTP}{GPRS Tunneling Protocol}
\newacronym{gtpc}{GTP-C}{GPRS Tunnelling Protocol Control Plane}
\newacronym{gtpu}{GTP-U}{GPRS Tunnelling Protocol User Plane}
\newacronym{gtpv2c}{GTPv2-C}{\gls{gtp} v2 - Control}
\newacronym{gw}{GW}{Gateway}
\newacronym{harq}{HARQ}{Hybrid Automatic Repeat reQuest}
\newacronym{hetnet}{HetNet}{Heterogeneous Network}
\newacronym{hh}{HH}{Hard Handover}
\newacronym{hol}{HOL}{Head-of-Line}
\newacronym{hqf}{HQF}{Highest-quality-first}
\newacronym{hss}{HSS}{Home Subscription Server}
\newacronym{http}{HTTP}{HyperText Transfer Protocol}
\newacronym{ia}{IA}{Initial Access}
\newacronym{iab}{IAB}{Integrated Access and Backhaul}
\newacronym{ic}{IC}{Incident Command}
\newacronym{ietf}{IETF}{Internet Engineering Task Force}
\newacronym{imsi}{IMSI}{International Mobile Subscriber Identity}
\newacronym{imt}{IMT}{International Mobile Telecommunication}
\newacronym{iot}{IoT}{Internet of Things}
\newacronym{ip}{IP}{Internet Protocol}
\newacronym{itu}{ITU}{International Telecommunication Union}
\newacronym{kpi}{KPI}{Key Performance Indicator}
\newacronym{kpm}{KPM}{Key Performance Measurement}
\newacronym{kvm}{KVM}{Kernel-based Virtual Machine}
\newacronym{los}{LOS}{Line-of-Sight}
\newacronym{lsm}{LSM}{Link-to-System Mapping}
\newacronym{lstm}{LSTM}{Long Short Term Memory}
\newacronym{lte}{LTE}{Long Term Evolution}
\newacronym{lxc}{LXC}{Linux Container}
\newacronym{m2m}{M2M}{Machine to Machine}
\newacronym{mac}{MAC}{Medium Access Control}
\newacronym{manet}{MANET}{Mobile Ad Hoc Network}
\newacronym{mano}{MANO}{Management and Orchestration}
\newacronym{mc}{MC}{Multi-Connectivity}
\newacronym{mcc}{MCC}{Mobile Cloud Computing}
\newacronym{mchem}{MCHEM}{Massive Channel Emulator}
\newacronym{mcs}{MCS}{Modulation and Coding Scheme}
\newacronym{mec}{MEC}{Multi-access Edge Computing}
\newacronym{mec2}{MEC}{Mobile Edge Cloud}
\newacronym{mfc}{MFC}{Mobile Fog Computing}
\newacronym{mgen}{MGEN}{Multi-Generator}
\newacronym{mi}{MI}{Mutual Information}
\newacronym{mib}{MIB}{Master Information Block}
\newacronym{miesm}{MIESM}{Mutual Information Based Effective SINR}
\newacronym{mimo}{MIMO}{Multiple Input, Multiple Output}
\newacronym{ml}{ML}{Machine Learning}
\newacronym{mlr}{MLR}{Maximum-local-rate}
\newacronym[plural=\gls{mme}s,firstplural=Mobility Management Entities (MMEs)]{mme}{MME}{Mobility Management Entity}
\newacronym{mmtc}{mMTC}{Massive Machine-Type Communications}
\newacronym{mmwave}{mmWave}{millimeter wave}
\newacronym{mpdccp}{MP-DCCP}{Multipath Datagram Congestion Control Protocol}
\newacronym{mptcp}{MPTCP}{Multipath TCP}
\newacronym{mr}{MR}{Maximum Rate}
\newacronym{mrdc}{MR-DC}{Multi \gls{rat} \gls{dc}}
\newacronym{mse}{MSE}{Mean Square Error}
\newacronym{mss}{MSS}{Maximum Segment Size}
\newacronym{mt}{MT}{Mobile Termination}
\newacronym{mtd}{MTD}{Machine-Type Device}
\newacronym{mtu}{MTU}{Maximum Transmission Unit}
\newacronym{mumimo}{MU-MIMO}{Multi-user \gls{mimo}}
\newacronym{mvno}{MVNO}{Mobile Virtual Network Operator}
\newacronym{nalu}{NALU}{Network Abstraction Layer Unit}
\newacronym{nas}{NAS}{Network Attached Storage}
\newacronym{nat}{NAT}{Network Address Translation}
\newacronym{nbiot}{NB-IoT}{Narrow Band IoT}
\newacronym{nfv}{NFV}{Network Function Virtualization}
\newacronym{nfvi}{NFVI}{Network Function Virtualization Infrastructure}
\newacronym{ni}{NI}{Network Interfaces}
\newacronym{nic}{NIC}{Network Interface Card}
\newacronym{nlos}{NLOS}{Non-Line-of-Sight}
\newacronym{now}{NOW}{Non Overlapping Window}
\newacronym{nsm}{NSM}{Network Service Mesh}
\newacronym{nrf}{NRF}{Network Repository Function}
\newacronym{nsa}{NSA}{Non Stand Alone}
\newacronym{nse}{NSE}{Network Slicing Engine}
\newacronym{nssf}{NSSF}{Network Slice Selection Function}
\newacronym{o2i}{O2I}{Outdoor to Indoor}
\newacronym{oai}{OAI}{OpenAirInterface}
\newacronym{oaicn}{OAI-CN}{\gls{oai} \acrlong{cn}}
\newacronym{oairan}{OAI-RAN}{\acrlong{oai} \acrlong{ran}}
\newacronym{oam}{OAM}{Operations, Administration and Maintenance}
\newacronym{ofdm}{OFDM}{Orthogonal Frequency Division Multiplexing}
\newacronym{olia}{OLIA}{Opportunistic Linked Increase Algorithm}
\newacronym{omec}{OMEC}{Open Mobile Evolved Core}
\newacronym{onap}{ONAP}{Open Network Automation Platform}
\newacronym{onf}{ONF}{Open Networking Foundation}
\newacronym{onos}{ONOS}{Open Networking Operating System}
\newacronym{oom}{OOM}{\gls{onap} Operations Manager}
\newacronym{opnfv}{OPNFV}{Open Platform for \gls{nfv}}
\newacronym{oran}{O-RAN}{Open Radio Access Network}
\newacronym{orbit}{ORBIT}{Open-Access Research Testbed for Next-Generation Wireless Networks}
\newacronym{os}{OS}{Operating System}
\newacronym{oss}{OSS}{Operations Support System}
\newacronym{pa}{PA}{Position-aware}
\newacronym{pase}{PASE}{Prioritization, Arbitration, and Self-adjusting Endpoints}
\newacronym{pawr}{PAWR}{Platforms for Advanced Wireless Research}
\newacronym{pbch}{PBCH}{Physical Broadcast Channel}
\newacronym{pcef}{PCEF}{Policy and Charging Enforcement Function}
\newacronym{pcfich}{PCFICH}{Physical Control Format Indicator Channel}
\newacronym{pcrf}{PCRF}{Policy and Charging Rules Function}
\newacronym{pdcch}{PDCCH}{Physical Downlink Control Channel}
\newacronym{pdcp}{PDCP}{Packet Data Convergence Protocol}
\newacronym{pdsch}{PDSCH}{Physical Downlink Shared Channel}
\newacronym{pdu}{PDU}{Packet Data Unit}
\newacronym{pf}{PF}{Proportional Fair}
\newacronym{pgw}{PGW}{Packet Gateway}
\newacronym{phich}{PHICH}{Physical Hybrid ARQ Indicator Channel}
\newacronym{phy}{PHY}{Physical}
\newacronym{pmch}{PMCH}{Physical Multicast Channel}
\newacronym{pmi}{PMI}{Precoding Matrix Indicators}
\newacronym{powder}{POWDER}{Platform for Open Wireless Data-driven Experimental Research}
\newacronym{ppo}{PPO}{Proximal Policy Optimization}
\newacronym{ppp}{PPP}{Poisson Point Process}
\newacronym{prach}{PRACH}{Physical Random Access Channel}
\newacronym{prb}{PRB}{Physical Resource Block}
\newacronym{psnr}{PSNR}{Peak Signal to Noise Ratio}
\newacronym{pss}{PSS}{Primary Synchronization Signal}
\newacronym{pucch}{PUCCH}{Physical Uplink Control Channel}
\newacronym{pusch}{PUSCH}{Physical Uplink Shared Channel}
\newacronym{qam}{QAM}{Quadrature Amplitude Modulation}
\newacronym{qci}{QCI}{\gls{qos} Class Identifier}
\newacronym{qoe}{QoE}{Quality of Experience}
\newacronym{qos}{QoS}{Quality of Service}
\newacronym{quic}{QUIC}{Quick UDP Internet Connections}
\newacronym{rach}{RACH}{Random Access Channel}
\newacronym[firstplural=Radio Access Technologies (RATs)]{rat}{RAT}{Radio Access Technology}
\newacronym{rbg}{RBG}{Resource Block Group}
\newacronym{rcn}{RCN}{Research Coordination Network}
\newacronym{rc}{RC}{RAN Control}
\newacronym{rec}{REC}{Radio Edge Cloud}
\newacronym{red}{RED}{Random Early Detection}
\newacronym{renew}{RENEW}{Reconfigurable Eco-system for Next-generation End-to-end Wireless}
\newacronym{rf}{RF}{Radio Frequency}
\newacronym{rfc}{RFC}{Request for Comments}
\newacronym{rfr}{RFR}{Random Forest Regressor}
\newacronym{ric}{RIC}{RAN Intelligent Controller}
\newacronym{rlc}{RLC}{Radio Link Control}
\newacronym{rlf}{RLF}{Radio Link Failure}
\newacronym{rlnc}{RLNC}{Random Linear Network Coding}
\newacronym{rmr}{RMR}{RIC Message Router}
\newacronym{rmse}{RMSE}{Root Mean Squared Error}
\newacronym{rnis}{RNIS}{Radio Network Information Service}
\newacronym{rr}{RR}{Round Robin}
\newacronym{rrc}{RRC}{Radio Resource Control}
\newacronym{rrm}{RRM}{Radio Resource Management}
\newacronym{rru}{RRU}{Remote Radio Unit}
\newacronym{rs}{RS}{Remote Server}
\newacronym{rsrp}{RSRP}{Reference Signal Received Power}
\newacronym{rsrq}{RSRQ}{Reference Signal Received Quality}
\newacronym{rss}{RSS}{Received Signal Strength}
\newacronym{rssi}{RSSI}{Received Signal Strength Indicator}
\newacronym{rtt}{RTT}{Round Trip Time}
\newacronym{ru}{RU}{Radio Unit}
\newacronym{rw}{RW}{Receive Window}
\newacronym{rx}{RX}{Receiver}
\newacronym{s1ap}{S1AP}{S1 Application Protocol}
\newacronym{sa}{SA}{standalone}
\newacronym{sack}{SACK}{Selective Acknowledgment}
\newacronym{sap}{SAP}{Service Access Point}
\newacronym{sc2}{SC2}{Spectrum Collaboration Challenge}
\newacronym{scef}{SCEF}{Service Capability Exposure Function}
\newacronym{sch}{SCH}{Secondary Cell Handover}
\newacronym{scoot}{SCOOT}{Split Cycle Offset Optimization Technique}
\newacronym{sctp}{SCTP}{Stream Control Transmission Protocol}
\newacronym{sdap}{SDAP}{Service Data Adaptation Protocol}
\newacronym{sdk}{SDK}{Software Development Kit}
\newacronym{sdm}{SDM}{Space Division Multiplexing}
\newacronym{sdma}{SDMA}{Spatial Division Multiple Access}
\newacronym{sdn}{SDN}{Software-defined Networking}
\newacronym[plural=SDRS]{sdr}{SDR}{Software-defined Radio}
\newacronym{seba}{SEBA}{SDN-Enabled Broadband Access}
\newacronym{sgsn}{SGSN}{Serving GPRS Support Node}
\newacronym{sgw}{SGW}{Service Gateway}
\newacronym{si}{SI}{Study Item}
\newacronym{sib}{SIB}{Secondary Information Block}
\newacronym{sinr}{SINR}{Signal to Interference plus Noise Ratio}
\newacronym{sip}{SIP}{Session Initiation Protocol}
\newacronym{siso}{SISO}{Single Input, Single Output}
\newacronym{sla}{SLA}{Service Level Agreement}
\newacronym{sm}{SM}{Service Model}
\newacronym{smf}{SMF}{Session Management Function}
\newacronym{smo}{SMO}{Service Management and Orchestration}
\newacronym{sms}{SMS}{Short Message Service}
\newacronym{smsgmsc}{SMS-GMSC}{\gls{sms}-Gateway}
\newacronym{snr}{SNR}{Signal-to-Noise-Ratio}
\newacronym{son}{SON}{Self-Organizing Network}
\newacronym{sptcp}{SPTCP}{Single Path TCP}
\newacronym{srb}{SRB}{Service Radio Bearer}
\newacronym{srn}{SRN}{Standard Radio Node}
\newacronym{srs}{SRS}{Sounding Reference Signal}
\newacronym{ss}{SS}{Synchronization Signal}
\newacronym{sss}{SSS}{Secondary Synchronization Signal}
\newacronym{st}{ST}{Spanning Tree}
\newacronym{svc}{SVC}{Scalable Video Coding}
\newacronym{tb}{TB}{Transport Block}
\newacronym{tcp}{TCP}{Transmission Control Protocol}
\newacronym{tdd}{TDD}{Time Division Duplexing}
\newacronym{tdm}{TDM}{Time Division Multiplexing}
\newacronym{tdma}{TDMA}{Time Division Multiple Access}
\newacronym{tfl}{TfL}{Transport for London}
\newacronym{tfrc}{TFRC}{TCP-Friendly Rate Control}
\newacronym{tft}{TFT}{Traffic Flow Template}
\newacronym{tgen}{TGEN}{Traffic Generator}
\newacronym{tip}{TIP}{Telecom Infra Project}
\newacronym{tm}{TM}{Transparent Mode}
\newacronym{to}{TO}{Telco Operator}
\newacronym{tr}{TR}{Technical Report}
\newacronym{trp}{TRP}{Transmitter Receiver Pair}
\newacronym{ts}{TS}{Technical Specification}
\newacronym{tti}{TTI}{Transmission Time Interval}
\newacronym{ttt}{TTT}{Time-to-Trigger}
\newacronym{tx}{TX}{Transmitter}
\newacronym{uas}{UAS}{Unmanned Aerial System}
\newacronym{uav}{UAV}{Unmanned Aerial Vehicle}
\newacronym{udm}{UDM}{Unified Data Management}
\newacronym{udp}{UDP}{User Datagram Protocol}
\newacronym{udr}{UDR}{Unified Data Repository}
\newacronym{ue}{UE}{User Equipment}
\newacronym{uhd}{UHD}{\gls{usrp} Hardware Driver}
\newacronym{ul}{UL}{Uplink}
\newacronym{um}{UM}{Unacknowledged Mode}
\newacronym{uml}{UML}{Unified Modeling Language}
\newacronym{upa}{UPA}{Uniform Planar Array}
\newacronym{upf}{UPF}{User Plane Function}
\newacronym{urllc}{URLLC}{Ultra Reliable and Low Latency Communications}
\newacronym{usa}{U.S.}{United States}
\newacronym{usim}{USIM}{Universal Subscriber Identity Module}
\newacronym{usrp}{USRP}{Universal Software Radio Peripheral}
\newacronym{utc}{UTC}{Urban Traffic Control}
\newacronym{vim}{VIM}{Virtualization Infrastructure Manager}
\newacronym{vm}{VM}{Virtual Machine}
\newacronym{vnf}{VNF}{Virtual Network Function}
\newacronym{volte}{VoLTE}{Voice over \gls{lte}}
\newacronym{voltha}{VOLTHA}{Virtual OLT HArdware Abstraction}
\newacronym{vr}{VR}{Virtual Reality}
\newacronym{vran}{vRAN}{Virtualized \gls{ran}}
\newacronym{vss}{VSS}{Video Streaming Server}
\newacronym{wbf}{WBF}{Wired Bias Function}
\newacronym{wf}{WF}{Waterfilling}
\newacronym{wg}{WG}{Working Group}
\newacronym{wlan}{WLAN}{Wireless Local Area Network}
\newacronym{osm}{OSM}{Open Source \gls{nfv} Management and Orchestration}
\newacronym{pnf}{PNF}{Physical Network Function}
\newacronym{drl}{DRL}{Deep Reinforcement Learning}
\newacronym{mtc}{MTC}{Machine-type Communications}
\newacronym{mns}{MnS}{Management Services}
\newacronym{ves}{VES}{\gls{vnf} Event Stream}
\newacronym{ei}{EI}{Enrichment Information}
\newacronym{fh}{FH}{Fronthaul}
\newacronym{fft}{FFT}{Fast Fourier Transform}
\newacronym{laa}{LAA}{Licensed-Assisted Access}
\newacronym{plfs}{PLFS}{Physical Layer Frequency Signals}
\newacronym{ptp}{PTP}{Precision Time Protocol}
\newacronym{cbrs}{CBRS}{Citizen Broadband Radio Service}

\newacronym{cif}{CI}{cyberinfrastructure}
\newacronym{sonic}{SONiC}{Software for Open Networking in the Cloud}
\newacronym{ocp}{OCP}{Open Compute Project}
\newacronym{snmp}{SNMP}{Simple Network Management Protocol}
\newacronym{raid}{RAID}{redundant array of independent disks}
\newacronym{nfs}{NFS}{Network File Storage}
\newacronym{ci}{CI}{Continuous Integration}
\newacronym{cd}{CD}{Continuous Deployment}
\newacronym{dtn}{DTN}{Data Transfer Node}

\newacronym{dns}{DNS}{Domain Name Service}
\newacronym{nrpe}{NRPE}{Nagios Remote Plugin Executor}
\newacronym{ldap}{LDAP}{Lightweight Directory Access Protocol}
\newacronym{lan}{LAN}{Local Area Network}
\newacronym{vlan}{VLAN}{Virtual LAN}

\newacronym{ipmi}{IPMI}{Intelligent Platform Management Interface}
\newacronym{tor}{ToR}{Top-of-the-Rack}
\newacronym{lmn}{LMN}{Large Memory Node}
\newacronym{bgp}{BGP}{Border Gateway Protocol}
\newacronym{dhcp}{DHCP}{Dynamic Host Configuration Protocol}
\newacronym{vrf}{VRF}{Virtual Routing and Forwarding}
\newacronym{vpn}{VPN}{Virtual Private Network}
\newacronym{rma}{RMA}{Return Merchandise Authorization}
\newacronym{hpc}{HPC}{High Performance Compute}

\newacronym{nu}{NU}{Northeastern University}
\newacronym{asic}{ASIC}{Application-specific Integrated Circuit}
\newacronym{rdma}{RDMA}{Remote Direct Memory Access}
\newacronym{roce}{RoCE}{RDMA over Converged Ethernet}
\newacronym{ovs}{OVS}{Open vSwitch}
\newacronym{frr}{FRR}{Free Range Routing}
\newacronym{ups}{UPS}{Uninterruptible Power Supply}

\newacronym{ntia}{NTIA}{National Telecommunications and Information Administration}
\newacronym{irb}{IRB}{Institutional Review Board}
\newacronym{doi}{DOI}{Digital Object Identifier}

\newacronym{sdo}{SDO}{Standard-Development Organization}
\newacronym{ose}{OSE}{Open Source Ecosystem}
\newacronym{osc}{OSC}{O-RAN Software Community}
\newacronym{dop}{DOP}{Director of Operations}
\newacronym{pm}{PM}{Program Manager}
\newacronym{excom}{EXCOM}{Executive Committee}
\newacronym{iiot}{IIoT}{Industrial \gls{iot}}
\newacronym{lf}{LF}{Linux Foundation}

\newacronym{wiot}{WIoT}{Institute for the Wireless Internet of Things}

\newacronym{otic}{OTIC}{Open Testing \& Integration Centre}

\newacronym{nofo}{NOFO}{Notice of Funding Opportunity}

\newacronym{onr}{ONR}{Office of Naval Research}
\newacronym{afosr}{AFOSR}{Air Force Office of Scientific Research}
\newacronym{afrl}{AFRL}{Air Force Research Laboratory}
\newacronym{arl}{ARL}{Army Research Laboratory}

\newacronym{arc}{ARC}{Aerial Research Cloud}

\newacronym{mno}{MNO}{Mobile Network Operator}
\newacronym{ct}{CT}{Continuous Testing}
\newacronym{oci}{OCI}{Open Container Initiative}

\newacronym[plural=RANs]{ran}{RAN}{Radio Access Network}
\newacronym{pii}{PII}{Personally Identifiable Information}
\newacronym{cves}{CVEs}{Common Vulnerabilities and Exposures}
\newacronym{cvss}{CVSS}{Common Vulnerability Scoring System}
\newacronym{n-rt-ric}{Near-RT RIC}{Near Real-Time RIC}
\newacronym{non-rt-ric}{Non-RT RIC}{Non Real-Time RIC}
\newacronym{o-cu}{O-CU}{O-RAN Central Unit}
\newacronym{o-cu-cp}{O-CU-CP}{\gls{o-cu} - Control Plane}
\newacronym{o-cu-up}{O-CU-UP}{\gls{o-cu} - User Plane}
\newacronym{o-du}{O-DU}{O-RAN Distributed Unit}
\newacronym{o-ru}{O-RU}{O-RAN Radio Unit}
\newacronym{sast}{SAST}{Static application security testing}
\newacronym{rbac}{RBAC}{Role-Based Access Control}
\newacronym{cis}{CIS}{Center for Internet Security}
\newacronym{ssrf}{SSRF}{Server-Side Request Forgery}
\newacronym[plural=NFs]{nf}{NF}{Network Function}

\newacronym{sub-manager}{SubM}{Subscription Manager}
\newacronym{dos}{DoS}{Denial of Service}
\newacronym{ac}{AC}{Attack Case}
\newacronym{cve}{CVE}{Common Vulnerabilities and Exposures}
\newacronym{cna}{CNA}{CVE Numbering Authority}
\newacronym{xapp}{xApp}{eXtended Application}
\newacronym{cwe}{CWE}{Common Weakness Enumeration}
\newacronym{rcmv}{RCMV}{Root Cause Mapping of Vulnerabilities}
\newacronym{e2ap}{E2AP}{E2 Application Protocol}
\newacronym{rest}{REST}{Representational State Transfer}
\newacronym[plural=$Sub_{Reqs}$]{sub-request}{$Sub_{Req}$}{Subscription Request}

\newacronym[plural=CAPECs]{capec}{CAPEC}{Common Attack Pattern Enumeration}
\newacronym[plural=APTs]{apt}{APT}{Advanced Persistent Threat}
\newacronym{attack}{MITRE ATT\&CK}{}
\newacronym{bsm}{BSM}{Base Score Metric}
\newacronym{fec}{FEC}{Forward Error Correction}
\newacronym{ldpc}{LDPC}{Low-Density Parity-Check}
\newacronym{adm}{ADM}{Accelerator Deployment Model}
\newacronym{aal}{AAL}{Acceleration Abstraction Layer}
\newacronym{bbdev}{BBDEV}{Base Band Device}
\newacronym{rc-d}{RC-D}{Regional Cloud}
\newacronym{ec-d}{EC-D}{Edge Cloud}
\newacronym{cs-d}{CS-D}{Cell Site}
\newacronym{cnf}{Cloudified NF}{Cloudified Network Function}
\newacronym{orkt}{ORKT}{O-RAN Key Technology}
\newacronym[plural=TTPs]{ttp}{TTP}{Tactics, Techniques and Procedure}
\newacronym{saas}{SaaS}{Software as a Service}
\newacronym{iaas}{IaaS}{Infrastructure as a Service}
\newacronym{ics-cert}{ICS-CERT}{Industrial Control Systems Cyber Emergency Response Team}
\newacronym{ics}{ICS}{Industrial Control Systems}
\newacronym{scada}{SCADA}{Supervisory Control and Data Acquisition}
\newacronym{json}{JSON}{JavaScript Object Notation}
\newacronym{mim}{MiM}{Man-in-the-Middle}
\newacronym[plural=DSPs]{dsp}{DSP}{Digital Signal Processing}
\newacronym[plural=DPDKs]{dpdk}{DPDK}{Data Plane Development Kit}
\newacronym{nvd}{NVD}{National Vulnerability Database}
\newacronym{nist}{NIST}{National Institute of Standards and Technology}

\newacronym{stix}{STIX}{Structured Threat Information Expression}
\newacronym{fcaps}{FCAPS}{Fault, Configuration, Accounting, Performance and Security}
\newacronym{cicd}{CI/CD}{Continuous Integration / Continuous Deployment}

\newacronym{yaml}{YAML}{Yet Another Markup Language}

\newacronym{csv}{CSV}{Comma-Separated Values}
\newacronym{hfc}{HFC}{Hard Filter Criterion}
\newacronym{sfc}{SFC}{Soft Filter Criterion}

\newacronym{fight}{MITRE FiGHT}{Framework for Intelligent Gathering and Harmonizing Threat intelligence}
\newacronym{poc}{PoC}{Proof-of-Concept}
\newacronym{nlp}{NLP}{Natural Language Processing}
\newacronym{tttm}{TTM}{Threat to Techniques-Mapping}
\newacronym{ttcm}{TCM}{Threat to CAPEC's-Mapping}
\newacronym{cuup}{CU-UP}{Centralized Unit - User Plane}
\newacronym{cucp}{CU-CP}{Centralized Unit - Control Plane}
\newacronym{dfd}{DFD}{Data Flow Diagram}

\pgfplotsset{/pgfplots/line legend 2/.style={
legend image code/.code={\draw[thick] (0cm,0cm)--(.25cm,0cm);},},}

\pgfplotsset{compat=1.11,
        /pgfplots/ybar legend/.style={
        /pgfplots/legend image code/.code={%
        \draw[##1,/tikz/.cd,bar width=3pt,yshift=-0.35em,bar shift=0pt]
                plot coordinates {(0cm,0.8em)};},
},
}

\newcommand{\vertcell}[2]{\rotatebox{90}{\begin{minipage}{#1}\begin{center}#2\end{center}\end{minipage}}}
\definecolor{critical}{RGB}{163, 28, 18}
\definecolor{critical_cvss}{RGB}{102, 15, 9}
\definecolor{high}{RGB}{238, 1, 5}
\definecolor{medium}{RGB}{241, 144, 1}
\definecolor{low}{RGB}{245, 206, 25}
\definecolor{negligible}{RGB}{78, 144, 255}

\usepackage{xspace}
\newcommand{\prop}{\texttt{ORCA}\xspace}

\newcommand*{\cicd}{\gls{ci}/\gls{cd}\xspace}
\newcommand*{\ct}{\gls{ct}\xspace}

\usepackage{listings}

\colorlet{punct}{red!60!black}
\definecolor{background}{HTML}{EEEEEE}
\definecolor{delim}{RGB}{20,105,176}
\colorlet{numb}{magenta!60!black}

\makeatletter
\def\lst@makecaption{%
  \def\@captype{table}%
  \@makecaption
}
\makeatother

\lstdefinelanguage{json}{
    basicstyle=\footnotesize\ttfamily,
    numbers=left,
    numberstyle=\scriptsize,
    stepnumber=1,
    numbersep=2pt,
    showstringspaces=false,
    breaklines=true,
    frame=lines,
    backgroundcolor=\color{background},
    literate=
     *{0}{{{\color{numb}0}}}{1}
      {1}{{{\color{numb}1}}}{1}
      {2}{{{\color{numb}2}}}{1}
      {3}{{{\color{numb}3}}}{1}
      {4}{{{\color{numb}4}}}{1}
      {5}{{{\color{numb}5}}}{1}
      {6}{{{\color{numb}6}}}{1}
      {7}{{{\color{numb}7}}}{1}
      {8}{{{\color{numb}8}}}{1}
      {9}{{{\color{numb}9}}}{1}
      {:}{{{\color{punct}{:}}}}{1}
      {,}{{{\color{punct}{,}}}}{1}
      {\{}{{{\color{delim}{\{}}}}{1}
      {\}}{{{\color{delim}{\}}}}}{1}
      {[}{{{\color{delim}{[}}}}{1}
      {]}{{{\color{delim}{]}}}}{1},
}

\def\BibTeX{{\rm B\kern-.05em{\sc i\kern-.025em b}\kern-.08em
    T\kern-.1667em\lower.7ex\hbox{E}\kern-.125emX}}
\usepackage{balance}
\begin{document}
\title{ORCA - An Automated Threat Analysis Pipeline\\ for O-RAN Continuous Development}
\author{\IEEEauthorblockN{Felix Klement\IEEEauthorrefmark{1}, Alessandro Brighente\IEEEauthorrefmark{2}, Michele Polese\IEEEauthorrefmark{3}, Mauro Conti\IEEEauthorrefmark{2}, Stefan Katzenbeisser\IEEEauthorrefmark{1}}\\
\IEEEauthorblockN{
\IEEEauthorrefmark{1}Computer Engineering, University of Passau, Passau, Germany\\
\IEEEauthorrefmark{2}Department of Mathematics, University of Padova, Padova, Italy\\
\IEEEauthorrefmark{3}Institute for the Wireless Internet of Things, Northeastern University, Boston, MA, USA\\
Email: \{felix.klement, stefan.katzenbeisser\}@uni-passau.de,\\\{alessandro.brighente, mauro.conti\}@unipd.it, m.polese@northeastern.edu
}

\IEEEaftertitletext{\vspace{-1\baselineskip}}


\thanks{This work was partially supported by the National Telecommunications and Information Administration (NTIA)'s Public Wireless Supply Chain Innovation Fund (PWSCIF) under Award No. 25-60-IF054. The authors acknowledge the financial support by the German
Federal Ministry of Research, Technologie and Aeronautics – Bundesministerium für Forschung, Technologie und Raumfahrt (BMFTR), as part of the Project “6G-RIC: The 6G Research
and Innovation Cluster” (project number 825026). Mauro Conti is also Wallenberg-WASP Guest Professor at Örebro University, Sweden}
}


\maketitle

\thispagestyle{empty}
\pagestyle{empty}

\begin{abstract}
The Open-Radio Access Network (O-RAN) integrates numerous software components in a cloud-like deployment, opening the radio access network to previously unconsidered security threats.
  With the ever-evolving threat landscape, integrating security practices through a DevSecOps approach is essential for fast and secure releases. Current vulnerability assessment practices often rely on manual, labor-intensive, and subjective investigations, leading to inconsistencies in the threat analysis. To mitigate these issues, we establish an automated pipeline that leverages Natural Language Processing (NLP) to minimize human intervention and associated biases. By mapping real-world vulnerabilities to predefined threat lists with a standardized input format, our approach is the first to enable iterative, quantitative, and efficient assessments, generating reliable threat scores for both individual vulnerabilities and entire system components within O-RAN. We illustrate the effectiveness of our framework through an example implementation for O-RAN, showcasing how continuous security testing can integrate into automated testing pipelines to address the unique security challenges of this paradigm shift in telecommunications.
\end{abstract}



\section{Introduction}
\gls{oran} represents a paradigm shift in the design of \gls{ran} for cellular networks~\cite{polese_understanding_oran}.
Thanks to its increased reliance on virtualization and softwarization rather than on proprietary and dedicated hardware components, \gls{oran} facilitates the development and integration of network functions similarly to more classical software supply chains.
The current pace of \gls{oran} innovation introduces a highly dynamic threat landscape that demands a close interaction between development and vulnerability assessment.
The attack surface in \gls{oran}, shaped by novel concepts, diverges significantly from conventional \glspl{ran}~\cite{LIYANAGE2023103621}. Some solutions already exist that can be used to tackle the new problems that arise in \gls{oran}. They must now be intelligently adapted and applied so that the security risks are eliminated or minimized.
Indeed, the increased transparency and accessibility represent an advantage to developers and attackers who can easily identify and exploit weak components.
Developing secure \gls{oran} demands solutions to automatically and efficiently assess, analyze, and monitor the threat surface. Numerous recently published \gls{oran} attacks and vulnerabilities~\cite{Mezzavilla2024DetectingOS, Ergu2024UnmaskingVA, Balakrishnan2024EnhancingOS, Chang2024PacketCD, Feliana2024EvaluationOC} underscore the necessity for an iterative and automated threat assessment framework ready to complement \gls{oran}'s continuous integration, delivery, and testing.

Shifting left security in software development is crucial for fast and secure code release and system deployment.
This is among the cornerstones of DevSecOps \cite{sanchez2020security}, where security testing shall be performed during development, testing, and integration.
To ease this process, a common approach is to scan software for known vulnerabilities and weaknesses via polling well-established public databases, such as the NIST \gls{nvd}.
Vulnerability detection is the first step of a security assessment: discovered vulnerabilities should be validated via a triage phase according to a budget-based priority list.
To this aim, tools such as \gls{capec} or MITRE ATT\&CK  provide a set of methodologies and attack patterns to guide triage.

\noindent{\textbf{Challenges in Threat Analysis.}}
A significant challenge for threat analysis is that it frequently remains a manual process.
As a result, the analysis and evaluation of complex systems is highly labor-intensive~\cite{RR-3188/2-AF}.
Additionally, it is often infeasible to rapidly replicate the analysis in response to emerging threat scenarios, new releases, or even small software updates.
Completing such assessments typically requires a substantial amount of personnel to avoid extending it over long time frames~\cite{s21144759}.
Another significant issue is the inherent subjectivity in manual analysis, as the outcome is influenced by the perspective and expertise of the individuals conducting the assessment.
This subjectivity can result in variations in the final results when the same analysis is performed multiple times, particularly when dealing with many threats.
Similarly, discrepancies may arise when different individuals conduct the analysis~\cite{Fakiha2023}.

There is no comprehensive solution for conducting automated assessments based on real vulnerabilities against a predefined generic threat list. Although numerous tools are available~\cite{Granata2023}, manual intervention is often required to address specific threats. Additionally, these tools are application-specific and lack flexibility for fine-tuning.

\noindent{\textbf{Contributions.}}
This paper proposes an automated threat analysis pipeline for \gls{oran} addressing the challenges mentioned above.
In particular, it eliminates manual evaluation by exclusively utilizing \gls{nlp}-based assignment.
This method minimizes the subjectivity typically associated with human decision-making.
Additionally, we assess and score threats based on potential sources of vulnerabilities through \glspl{cve}, rather than relying solely on subjective human judgment.
Utilizing a single input set of threat descriptions allows our analysis to be conducted iteratively and rapidly, consistently producing reliable results using the same input parameters.
Moreover, our framework enables the calculation of individual threat scores and the assessment of entire system components.
For instance, a heat map (as shown in Figure \ref{fig:tactics_headmap}) can be generated to identify specific MITRE tactic phases where improvements are necessary to enhance the security of the component.
This also serves as a valuable instrument for further evaluating identified weaknesses and developing potential solutions.
Beyond the mapping of the framework and the utilization of associated technique information, we develop a methodology specifically targeting existing \glspl{capec}.
This approach addresses gaps in the assessment areas of the MITRE frameworks, as not all techniques are necessarily linked to one or more \glspl{capec}.


To demonstrate our approach on the \gls{oran} scenario, we developed ORCA — \underline{O}pen-\underline{R}AN \underline{C}ontinuous Security \underline{A}nalysis based on our methodology. To enable a thorough and detailed assessment of the threats within \gls{oran}, we incorporate both the \gls{attack} and \gls{fight} frameworks. Including the complementary \gls{fight} framework during the preparation phase provides an opportunity to integrate additional security challenge details pertinent to telecommunication networks.

We summarize our contributions as follows.
\begin{itemize}
  \item We propose a novel \gls{nlp}-based methodology to map threat descriptions to standard tactics (ATT\&CK) and/or attack patterns (CAPEC) to reduce the inaccuracies of manual labor in quantifying risk scores. Our framework can be integrated into a live \gls{ci}/\gls{cd} pipeline and provide new results at every deploy or threat model update.
  \item We developed ORCA — Open-RAN Continuous Security Analysis, a realization of our pipeline on the \gls{oran} settings. We provide a detailed description of its realization to guide the development of our solution in a well-defined use case.
  \item We thoroughly test the performance of our approach through ORCA to assess the effectiveness and relevance of our proposal based on the threat definitions for the implementation of the O-RAN Software Community. We analyze the fine-tuning parameters we have introduced and our calculated scores for the set of generic threats.
\end{itemize}

\section{Related Work}
The domain of risk and threat analysis for system security has undergone extensive research historically. Prior studies~\cite{Hashim2018RiskAM, Kettani2019OnTT, Easttom2020ThreatA, Alsmadi2019CyberTA, Brkine2019NetworkTA} have employed the \gls{attack} framework, which will be utilized in this study as well. However, all of this work relates to classic risk, vulnerability, and threat analyses in computer science, some of which are very use-case specific and almost always have to be carried out manually. This is mainly due to context-specific factors, constantly evolving threats, and incomplete tool coverage. Thanks to the achievements in \gls{nlp}, these problems can now be overcome well, but integration into tools and frameworks is still lagging.

In the following, we provide a brief overview of three relevant key areas to our research: automated threat analysis, the automated mapping of components from the \gls{attack} framework to predefined threats, and, related to our use-case, the security evaluation of components within the \gls{oran} framework.

\noindent{\textbf{Automated Threat Research.}}
The field of automated threat modeling and analysis has received a lot of attention. However, most published work and tools still require manual steps at critical stages. This can distort the results and possibly falsify the analysis.
Granata et al.~\cite{Granata2023} analyzed three different graph-based methods for automating threat modeling. This included SLA-generator~\cite{Granata2021DesignAD}, Microsoft Theat tool~\cite{microsoft_threat_modeling} and Threat Dragon~\cite{threat_dragon_owasp}. They conclude that all the compared tools offer a more user-friendly approach than traditional methods and achieve comparably similar results. However, a major disadvantage is the limitation of the applicable domain within the tools.



\noindent{\textbf{Automated MITRE ATT\&CK Mapping.}}
Two noteworthy publications exist in automated mapping for the MITRE ATT\&CK framework. Both use \glspl{cve} as the basis for their machine-learning models.
Grigorescu et al.~\cite{Grigorescu2022CVE2ATTCKBM} propose a BERT-based mapping from \glspl{cve} to MITRE ATT\&CK techniques. The authors have annotated a data set of 1813 \glspl{cve} with the corresponding techniques. Hereby, they achieve an F1-score of 47.84\%.
Branescu et al.~\cite{info15040214} present a mapping from \glspl{cve} to MITRE ATT\&CK tactics. They experiment with transformer encoder architectures, encoder-decoder architectures, and zero-shot learning. The results show an F1 score of 78.88\% with consistent performance for the best-trained model.

\noindent{\textbf{Use-Case Related Threat Research.}}
An essential step towards the analysis of mobile communication systems was MITRE FiGHT \cite{mitre_fight}. Further work, such as that of Pell et al.~\cite{pell_mitre_moddeling_5G_core} attempts to improve protection against \glspl{apt} in 5G networks by closing the gaps in current 5G threat assessments. Wang et al.~\cite{wang_capec_cwe_6g} propose a methodology to build a detailed knowledge graph that enables the assessment of cyberattacks based on \glspl{capec} and \glspl{cwe} to support network intelligence in 6G networks.

Thimmaraju et al.~\cite{thimmaraju_oran} first investigate the supply chain risk for two large \gls{oran} \gls{n-rt-ric} implementations. They then present a run-time security testing tool for the A1 interface in the following part of their work. Their dependency analysis shows that \gls{oran} interfaces represent an opportunity for attackers to exploit the near-RT RIC. Mimran et al.~\cite{mimran_OPENRAN} present a security analysis for the architectural blueprint of the \gls{oran} Alliance. In their work, they have developed a taxonomy for the security analysis of \gls{oran}. This allows primary risk areas to be discovered, potential threats to be categorized, and various threat actors to be enumerated. Liyanage et al.~\cite{LIYANAGE2023103621} provide a comprehensive analysis of the security and privacy risks and challenges associated with the \gls{oran} architecture. They then discuss best practices that can be applied within the architecture.

Today, there is still no approach for a truly comprehensive assessment of all the threats defined in O-RAN. Furthermore, there is no standardized, repeatable methodology that allows for a stable assessment over time.

\section{ORCA:~Automated~Continuous~Security for O-RAN}
\begin{figure}[t]
  \centering
  \includegraphics[width=.8\linewidth]{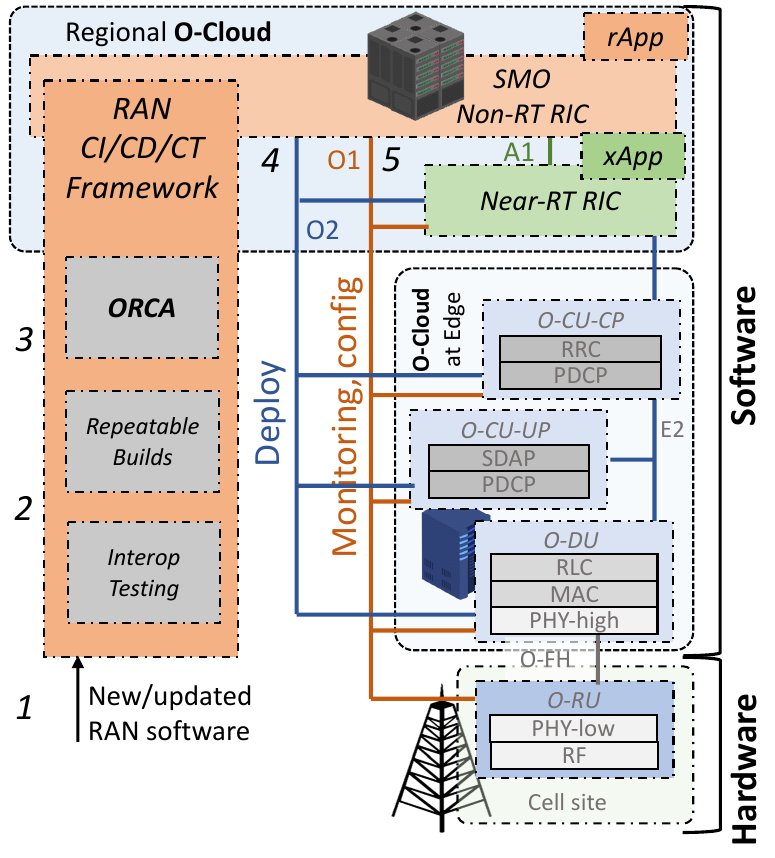}
  \caption{O-RAN architecture (right) and \prop as a component of \cicd and \ct pipelines for O-RAN.}
  \label{fig:o-ran}
\end{figure}

We demonstrate the capabilities of our pipeline approach in the context of an O-RAN system, by analyzing real-world software implementations and related threats for the complex, disaggregated elements that form a \gls{ran}. The specific instantiation of the threat score generation pipeline for O-RAN systems is called ORCA — \underline{O}pen-\underline{R}AN \underline{C}ontinuous Security \underline{A}nalysis.

A simplified logical diagram of the O-RAN architecture is presented in Fig.~\ref{fig:o-ran}, right. This architecture reflects the technical specifications defined by 3GPP and the O-RAN ALLIANCE, translating the principles of Open \gls{ran} systems into deployable solutions~\cite{o_ran_architecture_desc7}. These principles include: (i) an open system, characterized by standardized, open interfaces to foster a multi-vendor ecosystem; (ii) a disaggregated \gls{ran}, with functionalities distributed across different physical or virtual network functions; (iii) a software-driven approach, where components are deployed on white-box appliances and accelerators; and (iv) closed-loop control enabled by data-driven components deployed on \glspl{ric}~\cite{polese_understanding_oran}.

Figure~\ref{fig:o-ran} illustrates the disaggregated components of O-RAN~\cite{o_ran_architecture_desc7}.
Of these components, the \gls{ru} is typically implemented in hardware (using dedicated FPGAs or ASICs), while the remaining layers are implemented through software. Additional software components, such as the \glspl{ric} and \gls{smo}, provide control and orchestration functionalities connected via O-RAN interfaces (E2, O1, O2, A1) to the \gls{ran}~\cite{habibi2024unlocking}.

Software thus plays a key role in enabling next-generation cellular networks, offering flexibility, the ability to perform rolling upgrades, and full-stack programmability. However, the reliance on software also introduces challenges in testing, validating, and securing the complex stack~\cite{klement_jsac_oran,klement2024securing,dellOroRAN,brown2023heavy}. System integration and security assessments grow increasingly difficult as the stack's complexity expands. To address these challenges, \cicd and \ct have become integral components of O-RAN systems, as highlighted in~\cite{bonati20235gct,habibi2024unlocking}, and shown in Fig.~\ref{fig:o-ran}, left. Specifically, the O-Cloud, which designates the set of physical infrastructure and virtualization services supporting O-RAN systems, facilitates the automated deployment of new software functionalities through the O2 interface (in blue in Fig.~\ref{fig:o-ran}), managed by the \gls{smo}. The \gls{smo} also monitors the system via the O1 interface (in orange), detecting the status of network functions through keepalive messages, updating configurations, and collecting reports. Despite these advancements, integrating security elements into the system remains an open issue.

In this context, Fig.~\ref{fig:o-ran} also illustrates how \prop can be integrated into a \cicd and \ct pipeline for O-RAN, enabling continuous security evaluation. It primarily leverages the O2 interface, for deployment, and the O1 interface, for monitoring and configurations. These are exercised through a set of \cicd components in the \gls{smo}, according to the following workflow. \emph{First}, whenever there is new software to deploy (e.g., a new programmable application for the \glspl{ric}, called xApps and rApps, or a new implementation of a \gls{ran} network function), the operator submits the software package to the \cicd/\ct framework. \emph{Second,} this triggers a set of repeatable build pipelines~\cite{bonati20235gct} as well as tests that pertain to the functional domain and the interoperability with the rest of the system. \emph{Third,} in this context, we integrate \prop, which provides automated security testing without human intervention in the loop, as discussed in Sec.~\ref{sec:methodology}, making it a natural fit for the \cicd pipeline. \prop can assess the vulnerabilities in the new software, including their severity, mapping to known databases, and in case reject the tested element. \emph{Fourth}, if the tests pass, the software is deployed into the network using the O2 interface. \emph{Fifth}, post-deployment, the software is continuously monitored, and \prop can continue to assess the behavior of the software in real-time.

\section{Generating~Threat~Scores~from~Their Descriptions}
\label{sec:methodology}


This section presents our automated pipeline for generating threat scores from descriptions. 

\subsection{Pipeline Overview}\label{sec:overview}
\begin{figure}[!ht]
  \centering
  \includegraphics[width=.9\columnwidth]{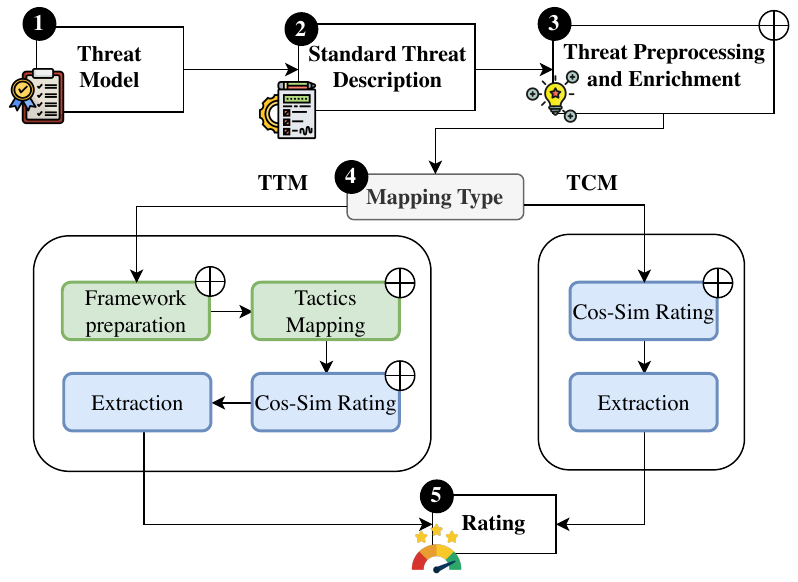}
  \caption{Simplified flow chart of our proposed pipeline.}
  \label{fig:pipeline}
\end{figure}

This section presents our \gls{oran} security analysis pipeline. Our approach can be used as a standalone tool or integrated into any existing \gls{ci}/\gls{cd} pipeline to check whether components or areas of interest are exposed to new vulnerabilities before the deployment goes live.
Figure~\ref{fig:pipeline} shows the simplified flow chart of the whole procedure. All steps indicated by a plus symbol are required only when modifications are made to the input threat dataset. If the threat data remains unchanged, only the extraction step needs to be performed, significantly reducing the overall execution time.

The initial step before running our approach involves threat modeling, a structured process for identifying, analyzing, and prioritizing potential security threats to an \gls{oran} system. Widely used tools and frameworks for threat modeling include the Microsoft Threat Modeling Tool \cite{microsoft_threat_modeling}, OWASP Threat Dragon \cite{threat_dragon_owasp}, and STRIDE \cite{stride}, all of which are designed to facilitate the identification and assessment of security risks in software and system environments. The modeled \gls{oran} threat data must subsequently be structured as a \gls{json} or \gls{csv} file, adhering to a pre-defined format that encapsulates detailed information about the individual threats. For each identified threat, we assume several key attributes, each represented in the dataset by a unique identifier. These attributes include a title, a description, an identified threat agent, a list of associated vulnerabilities, the relevant threatened resources within a system, and the affected components. A more comprehensive and precise characterization of these attributes enhances the quality of the final analysis. The initial values of the threat analysis are already sufficient for a successful use of ORCA. To ensure standardized input for subsequent processing within our pipeline, the threat information must be systematically organized. This structure can be customized to meet specific needs depending on the use case.

\subsection{Threat Preprocessing and Enrichment}
\label{subsec:data_preprocessing}
To provide the \gls{nlp} model with additional context information, such as the title and description of the threat, the data is integrated into the text paragraph that the model uses during preprocessing. In some cases, the relevant information may be distributed across multiple fields within the custom file format. To ensure the most accurate mapping, it is necessary to consolidate this distributed data into a standardized text paragraph. All of these processes are performed during the preprocessing step, which occurs once before the pipeline's execution. This step only needs to be repeated if the initial threat dataset changes. An example of such an approach is described in Section \ref{sub:oran_related_threat_preprocessing}.

\subsection{Mapping Type Branches}
Two different flow paths/branches can be executed individually or in parallel.
In our pipeline, we distinguish between: \gls{tttm} and \gls{ttcm}.
The former extracts the \gls{capec} information for the vulnerability analysis only from the assigned MITRE ATT\&CK techniques.
The latter accesses the \gls{capec} database directly and does not create any association with the MITRE frameworks. We present both methods because, in some cases, suitable mappings for specific threats may not be identified within the \gls{attack} techniques. The option to directly compare \glspl{capec} descriptions provides an additional approach to mitigate these omissions.

\subsection{TTM Branch}\label{sec:ttm}
\noindent{\textbf{Framework Preparation.}} The first major step in the \gls{tttm} branch is to extract, merge, and prepare the data from the \gls{attack} framework and merge it with additional complementary information. Such complementary information may be related to use-case-specific attack procedures (as shown in Section \ref{sub:orca_framework_preparations}) that guide our framework towards a more precise threat scoring.
This preprocessing step is conducted prior to the execution of our approach and does not impact the actual runtime of our pipeline.
The pre-extracted pieces of information are then mapped to the individual techniques described in the "Tactics Mapping" section below. To this aim, we leverage a sentence transformer to generate embeddings for threat descriptions and attack techniques. We then use cosine similarity to find \gls{tttm} mappings. Further information on this can be found in Section \ref{subsec:cos_sim} and details on the individual selection parameters used in the extraction step in Section \ref{subsec:fine_tuning}.


\noindent{\textbf{Tactics Mapping.}}
To automate the process of assigning relevant threats to techniques within \gls{attack} for extracting information from \glspl{capec}, \glspl{cwe}, \glspl{cve}, a number of steps are required. We realize them in an approach to find step-by-step the MITRE techniques with the greatest match.
In the following, we discuss the automated tactic mapping based on \cite{info15040214}. In this context, $R$ denotes the result set comprising potential mapped tactics derived from the outcomes of the individual language models. The merged set of all individual result sets is represented by $\psi$. The individual cosine similarities of the tactics within $\psi$ are quantified by $\mu$. Finally, $\xi$ represents the outcome corresponding to the maximum similarity between a tactic and a specific threat description.



To identify the potential matching techniques within the \gls{attack} framework used for each threat, we initially employ various models of \gls{nlp} to extract relevant tactics. The individual models, $\alpha - \delta$, sometimes generate distinct but valid outcomes, which are then combined. After eliminating duplicates, a consolidated set of suitable matches is produced.

\subsection{TCM Branch}
\label{sub:sub:capecs_branch}
In the \gls{ttcm} branch, we want an exact mapping between threats to \gls{capec} description.
Therefore, differently from \gls{tttm}, we do not include any additional complementary information.
The basis of this path is a dataframe of the latest \glspl{capec} database that is pulled for each \gls{ttcm} branch execution. The stages of Sections \ref{subsec:cos_sim} and \ref{subsec:fine_tuning} are then repeated analogously to the \gls{tttm} branch.



\subsection{Extraction and Rating}\label{sec:ear}

The extraction of the individual sets for techniques or \glspl{capec} are in themselves almost analogous with a few small differences. In general, however, the process can be described in a standardized way.  First, the created mapping files are read in, and a structured array with all the information from techniques or \glspl{capec} and threat id is created. With techniques mapping, an additional step is necessary here, as the \glspl{capec} of the techniques must first be determined. This foundation is subsequently utilized to identify additional relevant information based on the determined \glspl{capec} for each threat. The inclusion of further associated \glspl{capec} into the overall set at this stage of the pipeline, beyond those initially identified, depends on the selected scanning method. Each \gls{capec} is then iterated, and all associated \glspl{cwe} are extracted. In the final step, the \glspl{cve} corresponding to the identified \glspl{cwe} are identified, and their relevant metrics are recorded.
Starting from their description, we leverage standard indicators to assign severity scores to threats.

\subsection{\gls{oran} Related Threat Preprocessing}
\label{sub:oran_related_threat_preprocessing}
The threat information, which we extracted from the \gls{oran} Alliance's technical specification~\cite{wg11threatmodeling} (published as a Microsoft Word document), is grouped into eight fields per threat. In order to provide the sentence transformer, which calculates matching description paragraphs, with as much relevant information as possible, a single, comprehensible, continuous text must be generated from these fields. To enhance the contextual clarity of the individual entries from the threat input list for the \gls{nlp}, we utilize a short python program. This script appends contextual phrases, such as "A Threat with the title" or "and the description," before the title and description fields, respectively, to generate the complete body text.
This text paragraph is then the input for $x_1$ for matching the text equality in Section \ref{subsec:cos_sim}.

\subsection{ORCA Framework Preparation}
\label{sub:orca_framework_preparations}
The ORCA preprocessing is performed before the execution of the pipeline. For the enrichment, we utilize the publicly available FiGHT dataset, provided in the form of a \gls{yaml} file\footnote{https://github.com/mitre/FiGHT/blob/main/fight.yaml}.
FiGHT provides security information specific to a cellular network scenario, thus being a specialized version of ATT\&CK.
Unfortunately, not all techniques within FiGHT are provided with associated \glspl{capec}.
This means that no useful information can be extracted for our further threat rating calculation.
To solve this issue, we include an additional preprocessing step to obtain all useful metrics about potentially associated vulnerabilities that can be used.

Many of the techniques included in FiGHT have a 1:1 mapping to those in ATT\&CK, but unfortunately, not all. 51 out of 70 main techniques have no associated mapping or no addenda assigned.
We exclude all techniques in FiGHT that do not correspond to a matching ATT\&CK technique or lack supplementary materials or updates that extend the core framework. No relevant information can be derived from these filtered techniques for our subsequent analysis. This is because our evaluation relies on metrics that can be extracted using \glspl{capec}.
Subsequently, the additional information in FiGHT, the so-called addendums, are extracted from the remaining techniques.
An addendum is additional information in the context of the technique concerning 5G.
In addition, a technique can have several addenda to provide extensive information content in the 5G telecommunications setting.
The addenda, which we extracted from the \gls{yaml} file for the associated techniques, are then added in an additional column to enable even more specific threat-to-technique mapping in the telecommunications context.

\section{Implementation}
The implementation details of our approach are described in the following sections. An exemplary demonstration of how the pipeline's execution proceeds is provided in Appendix \ref{app:sec:example}.

\subsection{Tactics Mapping}

Branescu et al. \cite{info15040214} use a large \gls{cve} dataset with assigned tactic labels. Since their trained models are not publicly available, we retrained six of the eight models for our pipeline, following their data and methods. Table \ref{tab:lang_models} shows the performance of our re-trained models. While the tactic mapping in our approach is interchangeable, Branescu et al.'s approach currently offers the best operational quality.




\begin{table}[ht]
  \caption{Performance results for the dataset of each of the language models we re-trained based on \cite{info15040214}.}
  \resizebox{\columnwidth}{!}{%
    \begin{tabular}{ccc}
      \hline
      Model                                                                                     & Weighed F1 (SD)  & F1 Macro (SD)    \\ \hline & \\
      CyBERT \cite{9671824}                                                                     & 78.67\% (1.08\%) & 58.20\% (4.52\%) \\
      SecBERT \cite{aghaei2022securebert}                                                       & 79.96\% (0.43\%) & 68.07\% (2.25\%) \\
      SecRoBERTa \cite{liu2019roberta,Jackaduma_secroberta}                                     & 79.28\% (0.72\%) & 66.94\% (1.89\%) \\[0.2cm] \hline & \\
      \begin{tabular}[c]{@{}c@{}}TARS short \\ labels \cite{halder-etal-2020-task}\end{tabular} & 78.78\% (0.33\%) & 67.92\% (1.22\%) \\
      \begin{tabular}[c]{@{}c@{}}TARS long \\ labels \cite{halder-etal-2020-task}\end{tabular}  & 78,81\% (2.06\%) & 67,16\% (4.61\%) \\[0.2cm] \\ \hline
    \end{tabular}%
  }
  \label{tab:lang_models}
\end{table}

The trained language models are applied to the respective prepared dataset. The individual results, referred to as $R_\alpha$ to $R_\delta$, are stored in the set $\psi$, with the best matches finally being summarized in $\xi$. This set then serves as input for the subsequent stage of the pipeline, which is discussed in Section \ref{subsec:cos_sim}.


\subsection{Cosine Similarity Rating}
\label{subsec:cos_sim}


We use SBERT \cite{reimers-2019-sentence-bert} to generate embeddings from preprocessed strings. Threat descriptions often differ in wording but convey similar ideas. SBERT captures these semantic relationships, unlike keyword-based methods (e.g., TF-IDF) or simple embeddings (e.g., Word2Vec), which struggle with deeper meanings. It maps fine-grained differences and high-level similarities, aligning specific attack vectors with generalized patterns in techniques or \gls{capec} descriptions. SBERT efficiently embeds text in a vector space, where similar texts are closer and can be found using cosine similarity.


We use all-MiniLM-L12-v2\footnote{https://huggingface.co/sentence-transformers/all-MiniLM-L12-v2}, an all-round model trained on over a billion pairs for diverse use cases. It effectively understands cybersecurity-specific descriptions without requiring extensive fine-tuning. The model maps text into 384-dimensional vectors for clustering or semantic comparisons and is both fast and lightweight. We use cosine similarity to match \gls{oran} threats with extracted techniques from preselected tactics. While these models are interchangeable, evaluating NLP techniques for cybersecurity is beyond our scope. Preliminary work has already proven such feasibility in this subject area \cite{Shahid2021CVSSBERTEN}.





\subsubsection{Techniques Mapping}
We use our prepared text paragraph from \ref{subsec:data_preprocessing} with the detailed and comprehensive threat description as input for $x_1$. The techniques description from Section \ref{sec:ttm} with the added addendums serves as the input for $x_2$.
We then iterate all the techniques of the preselected tactics from $\xi$ for $x_2$. Using only a subset of tactics for the vector space matching from threat to techniques speeds up our approach many times over.

\subsubsection{CAPEC's Mapping}
As outlined in \ref{sub:sub:capecs_branch}, the \glspl{capec} mapping process does not necessitate any specially prepared input. Similar to techniques mapping, $x_1$ represents the extracted and pre-processed text paragraph, while $x_2$ corresponds to the description of the \glspl{capec}, which can be retrieved and accessed through \gls{stix} utilizing the dataset provided by MITRE. Each description is systematically iterated over, and the respective scores are recorded.

\subsection{Fine Tuning the Extraction}
\label{subsec:fine_tuning}

The results of a single batch run depend on a large number of parameters, which also have a considerable influence on the outcome and the total runtime of the pipeline. In the following, we discuss the most influential parameters.

\subsubsection{Threshold Definition}
To set how high the minimum equality between $x_1$ and $x_2$ should be in order to be included in the final set of techniques or \glspl{capec} for a threat, we define the threshold $\cos_{thrs}$. The equality value between the two paragraphs ranges from -1 to 1, where 1 means that the vectors are identical, 0 means that they are orthogonal to each other, and -1 means that they are exactly opposite. In principle, cosine similarity is a useful tool, but it may have some drawbacks. For example, different expressions with the same meaning can be represented by different vectors, which can lead to a lower equality value. Model dependency also plays a major role. Moreover, models trained on large and more diverse datasets tend to generate better semantic embeddings. The threshold serves as a heuristic to determine which techniques, among the extracted matching tactics, are better suited to address a given threat compared to others.

\subsubsection{Scanning Method}
\label{sec:scanningmethod}
We develop a function called \textit{deep\_scan} to extract not only the \gls{capec} associated \glspl{cve} mentioned, but also potentially related ones. We proceed in such a way that not only the individual CAPEC associated with the technology is taken into account, but also the relationships with \gls{stix} pattern. Specifically, we use the parent attribute \textit{x\_capec\_parent\_of\_refs}, which provides us with the associated child attack patterns and \textit{x\_capec\_can\_precede\_refs} that provides all attack patterns where the current underlying one may have been preceded.
The search is performed recursively.
In addition, we add flags that control for which attributes the recursion should also be performed.
We only perform a recursive search for the parents attribute to avoid extracting too many \glspl{capec}. This enables us to conduct searches for extractions either in the Normal-Mode (N), without incorporating additional potentially dependent \glspl{capec}, or in the previously described Deep-Mode (D).

\subsubsection{Filter Criteria}
To enable an analysis for each threat, we implement two filtering strategies: \gls{hfc} and \gls{sfc}.
For \gls{sfc}, the highest possible match of the cosine equality is included in the final dataset if no hit is above the threshold.
The situation is different for \gls{hfc}, where results are only included if the cosine equality is above the threshold.
However, this sometimes means that some threats cannot be evaluated because no \glspl{capec} or techniques can be assigned.
Depending on whether we want to have a potential hit included for all threats or not, we should carry out the analysis using \gls{sfc}. This way, at least one available match is assigned for each threat.
However, it can still happen that a threat cannot be assigned a score.
In some cases, mapped techniques or \glspl{capec} have no associated \glspl{cve}.

\subsubsection{Extraction Uniquness}
The uniqueness of \glspl{capec} and \glspl{cve} can also be enforced within the dataset through the boolean parameter $\omega$, which eliminates duplicate entries.
This option is introduced to mitigate excessively large extractions.
However, enabling uniqueness results in the loss of the feature where multiple occurrences of \glspl{capec} and \glspl{cve} contribute to the overall score.
When $\omega = \text{false}$, the scores from multiple occurrences are cumulative, meaning that, for instance, more frequent and severe \glspl{cve} will have a greater impact on the overall score than when $\omega = \text{true}$.

\subsubsection{Inclusion Period}
Another setting option is the selection of the date from which \glspl{cve} should be extracted or considered.
Here, $\tau$ specifies the date from which a \gls{cve} should be considered in our extraction.
With $\tau =$ 2024-01-01, only \glspl{cve} from the current year 2024 and later are included.
This makes it possible to exclude older vulnerabilities from the outset.
What is of greater advantage, however, is the fact that it allows the extraction to be carried out only from a specific last point in time, e.g., the date of the last scan.
This provides the rating for new \glspl{cve} found since that point in time and can thus specifically examine the new threat situation that has arisen since that point in time.
We use $\tau =$ 1998-01-01 to extract all currently existing \glspl{cve}.

In order to assess the efficiency of the individual parameters, we analyze various effects in Section \ref{subsec:benchmarking}. Specifically, we investigate how the result sets for both mapping strategies are affected by varying thresholds when altering the scan method and filter criteria. Additionally, we compare the two mapping strategies, \gls{tttm} and \gls{ttcm}, evaluate the validity of these approaches, and examine the impact of duplicates within the result set on the final calculated scores.


\subsection{Calculation of Score Metrics}
We utilize the widely adopted \gls{bsm} vector to compute metric scores. In our approach, the calculation of individual scores begins with selecting a specific version of the \gls{cvss}: version 2, 3, or 4. Each version is valid in its application and presents distinct advantages and limitations. The choice of version should be tailored to the specific requirement and applied consistently throughout the assessment. Additional classification values, such as Temporal Score Metrics, Supplemental Metrics, Environmental Score Metrics, and Threat Metrics, are frequently unavailable for many \glspl{cve}. Moreover, the specifications of the \gls{cvss} versions differ and do not consistently include the same additional metrics beyond the base score. Consequently, we restrict our analysis to the \gls{bsm}.

\section{Results}
In this section, we evaluate the parameters of the implemented pipeline, analyzing the impact of various thresholds on the scoring outcomes and comparing the effectiveness of the two mapping strategies employed.
Next, we examine the \glspl{cvss} scores generated for the list of general \gls{oran} threats and compare our exemplary results with those previously reported by the \gls{oran} Alliance.
Finally, we provide an outlook on potential future directions.

\subsection{Benchmarking of Tuning Parameters}
\label{subsec:benchmarking}
To quantify the variability of outcomes, we compare results across a realistic range of parameter settings.
Based on this comparison, we set the configuration for our analysis of \gls{oran} threats.

\subsubsection{Result Sets for Different Thresholds}
\label{subsec:result_sets_for_different_thresholds}
An important point at the beginning of the evaluation is the choice of the correct pipeline parameters and the mapping strategy.
Therefore, we analyzed three different realistic values from $0.45$ to $0.55$ in steps of $0.05$ for $\cos_{thrs}$ for the threat to techniques mapping and for the threat to \glspl{capec} mapping. Based on our experience, starting values of $0.45$ have consistently yielded reliable results. Whereby, threshold values exceeding $0.55$ tend to be overly restrictive, leading to the exclusion of many relevant mappings. Table~\ref{tab:overview_extraction_numbers} shows the number of \glspl{cve} and \glspl{capec} found for different initial values of $\cos_{thrs}$, for both scan modes, both filter criteria, and also the two mapping constellations.

\begingroup
\centering
\setlength{\tabcolsep}{7pt} 
\renewcommand{\arraystretch}{1.5} 
\begin{table*}
  \centering
  \caption{Overall number of extracted techniques, \glspl{capec}, \glspl{cve} with Deep-Mode and with Normal-Mode per specific threshold and filter criterion for both mapping strategies from the \gls{oran} general threats with $\omega = true$.}
  \label{tab:overview_extraction_numbers}
  \resizebox{\textwidth}{!}{
    \begin{tabular}{cccccccccccc}
                                                 &                                    & \multicolumn{5}{c}{Threat to Techniques-Mapping} & \multicolumn{5}{c}{Threat to CAPEC's-Mapping}                                                                                                                                                                                                                                                                         \\ \cmidrule(lr){3-7} \cmidrule(lr){8-12} \hline
      \multicolumn{1}{|c|}{\multirow{6}{*}{HFC}} & \multicolumn{1}{c|}{$\cos_{thrs}$} & \multicolumn{1}{c|}{Techniques}                  & \multicolumn{1}{c|}{CAPEC's}                  & \multicolumn{1}{c|}{DEEP\_CAPEC's} & \multicolumn{1}{c|}{CVE's} & \multicolumn{1}{c|}{DEEP\_CVE's} & \multicolumn{1}{c|}{Threats} & \multicolumn{1}{c|}{CAPEC's} & \multicolumn{1}{c|}{DEEP\_CAPEC's} & \multicolumn{1}{c|}{CVE's} & \multicolumn{1}{c|}{DEEP\_CVE's} \\ \cline{2-12}
      \multicolumn{1}{|c|}{}                     & \multicolumn{1}{c|}{0.45}          & \multicolumn{1}{c|}{13}                          & \multicolumn{1}{c|}{16}                       & \multicolumn{1}{c|}{58}            & \multicolumn{1}{c|}{10515} & \multicolumn{1}{c|}{41763}       & \multicolumn{1}{c|}{32}      & \multicolumn{1}{c|}{325}     & \multicolumn{1}{c|}{403}           & \multicolumn{1}{c|}{75322} & \multicolumn{1}{c|}{76068}       \\ \cline{2-12}
      \multicolumn{1}{|c|}{}                     & \multicolumn{1}{c|}{0.50}          & \multicolumn{1}{c|}{8}                           & \multicolumn{1}{c|}{11}                       & \multicolumn{1}{c|}{54}            & \multicolumn{1}{c|}{10078} & \multicolumn{1}{c|}{41482}       & \multicolumn{1}{c|}{29}      & \multicolumn{1}{c|}{193}     & \multicolumn{1}{c|}{319}           & \multicolumn{1}{c|}{51002} & \multicolumn{1}{c|}{72788}       \\ \cline{2-12}
      \multicolumn{1}{|c|}{}                     & \multicolumn{1}{c|}{0.55}          & \multicolumn{1}{c|}{6}                           & \multicolumn{1}{c|}{9}                        & \multicolumn{1}{c|}{52}            & \multicolumn{1}{c|}{8770}  & \multicolumn{1}{c|}{41482}       & \multicolumn{1}{c|}{18}      & \multicolumn{1}{c|}{69}      & \multicolumn{1}{c|}{174}           & \multicolumn{1}{c|}{41058} & \multicolumn{1}{c|}{60552}       \\ \hline \hline
      \multicolumn{1}{|c|}{\multirow{6}{*}{SFC}} & \multicolumn{1}{c|}{$\cos_{thrs}$} & \multicolumn{1}{c|}{Techniques}                  & \multicolumn{1}{c|}{CAPEC's}                  & \multicolumn{1}{c|}{DEEP\_CAPEC's} & \multicolumn{1}{c|}{CVE's} & \multicolumn{1}{c|}{DEEP\_CVE's} & \multicolumn{1}{c|}{Threats} & \multicolumn{1}{c|}{CAPEC's} & \multicolumn{1}{c|}{DEEP\_CAPEC's} & \multicolumn{1}{c|}{CVE's} & \multicolumn{1}{c|}{DEEP\_CVE's} \\ \cline{2-12}
      \multicolumn{1}{|c|}{}                     & \multicolumn{1}{c|}{0.45}          & \multicolumn{1}{c|}{13}                          & \multicolumn{1}{c|}{16}                       & \multicolumn{1}{c|}{58}            & \multicolumn{1}{c|}{10515} & \multicolumn{1}{c|}{41763}       & \multicolumn{1}{c|}{32}      & \multicolumn{1}{c|}{325}     & \multicolumn{1}{c|}{403}           & \multicolumn{1}{c|}{75322} & \multicolumn{1}{c|}{76068}       \\ \cline{2-12}
      \multicolumn{1}{|c|}{}                     & \multicolumn{1}{c|}{0.50}          & \multicolumn{1}{c|}{8}                           & \multicolumn{1}{c|}{11}                       & \multicolumn{1}{c|}{54}            & \multicolumn{1}{c|}{10078} & \multicolumn{1}{c|}{41482}       & \multicolumn{1}{c|}{32}      & \multicolumn{1}{c|}{196}     & \multicolumn{1}{c|}{320}           & \multicolumn{1}{c|}{51002} & \multicolumn{1}{c|}{72788}       \\ \cline{2-12}
      \multicolumn{1}{|c|}{}                     & \multicolumn{1}{c|}{0.55}          & \multicolumn{1}{c|}{8}                           & \multicolumn{1}{c|}{11}                       & \multicolumn{1}{c|}{54}            & \multicolumn{1}{c|}{10078} & \multicolumn{1}{c|}{41482}       & \multicolumn{1}{c|}{30}      & \multicolumn{1}{c|}{87}      & \multicolumn{1}{c|}{218}           & \multicolumn{1}{c|}{43281} & \multicolumn{1}{c|}{65806}       \\ \hline
    \end{tabular}
  }


\end{table*}
\endgroup


Figure~\ref{fig:score_matrix} shows the respective average \gls{cvss} v2 values for exploitability, base score and impact for each extracted cosine threshold value for the techniques mapping in the \gls{hfc}.
\begingroup
\centering
\setlength{\tabcolsep}{7pt} 
\renewcommand{\arraystretch}{1.5} 
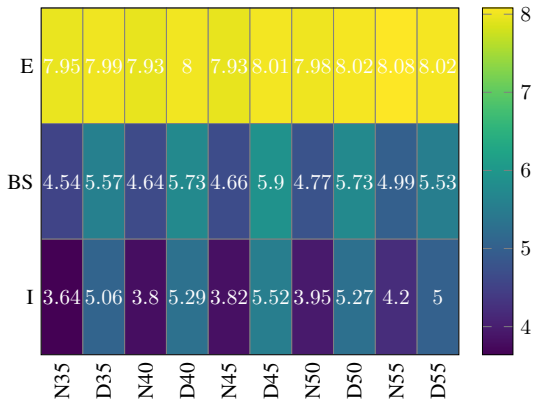
\begin{figure}[!ht]
  \begin{center}
    \vspace{1em}
    E $\equiv$ Exploitability, BS $\equiv$ Base Score, I $\equiv$ Impact \color{Black}\\
    N $\equiv$ Normal-Mode, D $\equiv$ Deep-Mode\\
    NXX $\vee$ DXX $\rightarrow$ XX $\equiv$ Percentage for $\cos_{thrs}$
  \end{center}
  \centering
  \begin{adjustbox}{width=0.4\textwidth}
    \begin{tikzpicture}
    \begin{axis}[
            colormap/viridis,
            xticklabels={N35, D35, N40, D40, N45, D45, N50, D50, N55, D55}, 
            xtick={0,...,9}, 
            xtick style={draw=none},
            yticklabels={E, BS, I}, 
            ytick={0,...,4}, 
            ytick style={draw=none},
            enlargelimits=false,
            colorbar,
            xticklabel style={
              rotate=90
            },
            nodes near coords={\pgfmathprintnumber\pgfplotspointmeta},
            nodes near coords style={
                yshift=-7pt
            },
        ]
        \addplot[
            matrix plot,
            mesh/cols=10, 
            point meta=explicit,draw=gray,text=white
        ] table [meta=C] {
            x y C 
            0 0 7.95 
            1 0 7.99
            2 0 7.93
            3 0 8.00
            4 0 7.93
            5 0 8.01
            6 0 7.98
            7 0 8.02
            8 0 8.08
            9 0 8.02
            0 1 4.54
            1 1 5.57
            2 1 4.64
            3 1 5.73
            4 1 4.66
            5 1 5.90
            6 1 4.77
            7 1 5.73
            8 1 4.99
            9 1 5.53
            0 2 3.64
            1 2 5.06
            2 2 3.80
            3 2 5.29
            4 2 3.82
            5 2 5.52
            6 2 3.95
            7 2 5.27
            8 2 4.20
            9 2 5.00

        }; 
    \end{axis}
\end{tikzpicture}
  \end{adjustbox}
  \caption{Calculated average \gls{cvss}v2 \gls{bsm} of the generic \gls{oran} threats for the technique mapping strategy using \gls{hfc}, $\omega = false$.}
  \label{fig:score_matrix}
\end{figure}
\endgroup

We note that the scores for the set of \glspl{cve} used for each of the three categories remain relatively stable throughout the different $\cos_{thrs}$.
For example, the coefficient of variation for exploitability is only $0.005$, for the base score $0.097$, and for the impact $0.153$.
This indicates that the data in our setting is relatively homogeneous and that the scatter is not very strong.
We conclude, that due to the similar basic sets of \glspl{cve}, a meaningful overall assessment of the system can also be made by averaging over a smaller set.
For this assessment, a larger number of total \glspl{cve} is therefore not necessary for an averaged overall result, as long as the significant basic vulnerabilities are identified reliably.
However, if a detailed and extensive analysis is desired as opposed to a quick and brief assessment, lower values for the cosine equality must definitely be used.
This covers as many potential security problems as possible that have a certain equality or connection to the predefined threats.

\subsubsection{Comparison of Mapping Strategies}
To investigate discrepancies between the two mapping strategies \gls{tttm} and \gls{ttcm}, we calculated the average values for impact, exploitability, and the base score in normal mode for the two $\cos_{thrs}$ of $0.50$ and $0.55$ under the \gls{hfc} filter criterion.
The individual values of the respective scores can be found in Figure \ref{fig:mapping_strategies_comp}.
At first glance, a close clustering around the average values of the individual score classes for the exploitability can be seen, only for the impact and the base score, the deviation is slightly larger.
The average coefficient of variation of the individual scores is $0.08$, which we consider acceptable.
From this, we derive that both mapping strategies can achieve meaningful results with only slight nuances.

\begingroup
\centering
\setlength{\tabcolsep}{7pt} 
\renewcommand{\arraystretch}{1.5} 
\begin{figure}[H]
  \centering
  \begin{adjustbox}{width=0.4\textwidth}
    \begin{tikzpicture}

\definecolor{green}{RGB}{0,127,0}
\definecolor{amber}{rgb}{1.0, 0.75, 0.0}
\definecolor{bleudefrance}{rgb}{0.19, 0.55, 0.91}
\definecolor{darkcoral}{rgb}{0.8, 0.36, 0.27}
\definecolor{darkcyan}{rgb}{0.0, 0.55, 0.55}
\definecolor{crimsonglory}{rgb}{0.75, 0.0, 0.2}

\tikzstyle{every node}=[font=\footnotesize]

\begin{axis}[
    ybar,
    ymin=0,
    ymax=10,
    ymajorgrids = true,
    minor y tick num=6,
    ylabel={CVSSv2 Score},
    enlarge x limits=0.28,
    legend style={at={(0.5,-0.15)},
      anchor=north,legend columns=-1},
    symbolic x coords={Impact,Exploitability,Base Score},
    xtick=data
    ]
\addplot [black, fill=amber,postaction={pattern=north west lines},nodes near coords,
    every node near coord/.append style={font=\footnotesize},
    nodes near coords align={vertical}] coordinates {(Impact,4.0) (Exploitability,8.0) (Base Score,4.8)};
\addplot [black,fill=bleudefrance,postaction={pattern=north east lines},nodes near coords,
    every node near coord/.append style={font=\footnotesize},
    nodes near coords align={vertical}] coordinates {(Impact,4.2) (Exploitability,8.1) (Base Score,5.0)};
\addplot [black,fill=darkcoral,postaction={pattern=crosshatch},nodes near coords,
    every node near coord/.append style={font=\footnotesize},
    nodes near coords align={vertical},] coordinates {(Impact,5.6) (Exploitability,8.0) (Base Score,6.0)};
\addplot [black,fill=darkcyan,postaction={pattern=grid},nodes near coords,
    every node near coord/.append style={font=\footnotesize},
    nodes near coords align={vertical},] coordinates {(Impact,5.6) (Exploitability,8.0) (Base Score,6.0)};

\addplot[draw=crimsonglory,ultra thick,smooth,mark=*] 
    coordinates {(Impact,4.85) (Exploitability,8.025) (Base Score, 5.45)};

\legend{TTM-N050,TTM-N055,TCM-N050, TCM-N055, Avg. Scoring}
\end{axis}
\end{tikzpicture}
  \end{adjustbox}
  \caption{Average \gls{cvss}v2 scores for $\cos_{thrs}$ 0.50 and 0.55 with $\omega = false$ using the \gls{hfc} criterion with Normal-Mode for \gls{tttm} and \gls{ttcm}.}
  \label{fig:mapping_strategies_comp}
\end{figure}
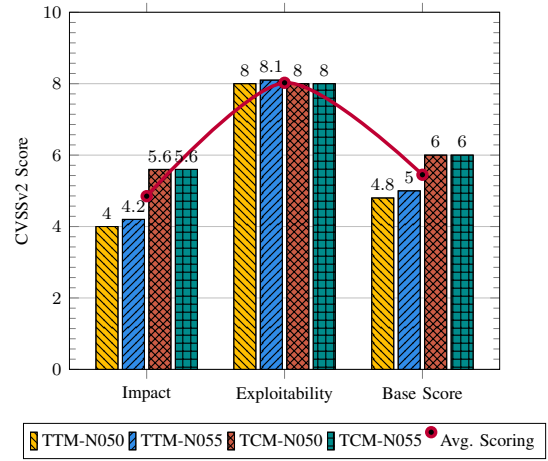
\endgroup

\subsubsection{Validity of Mapping Strategies}
To check the accuracy and reliability of the technique mapping, we carried out manual spot checks.
To do this, we extracted a random set of mappings multiple times and checked whether they matched each other correctly. Taking multiple random samples ensures that different parts of the mapping set are checked, which increases reliability.
We examined $3$ rounds each with a total of $60$ assignments. The average percentage of non-matching mappings is $5.5 \%$. With 235 technique mappings, there is a theoretical coverage of 76.6\% if each assignment is checked once. Since the 60 mappings per round were randomly selected, most mappings were likely checked at least once, and some even multiple times. This approach reduces bias and increases the reliability of the results.

We did the same for the \glspl{capec} mapping strategy (with a total of 316 mappings) and achieved $3.9 \%$ mismatches in our $3$ runs with $60$ assignments.
We explain the slightly better agreement with the greater diversity exhibited by the number of \glspl{capec} ($594$ in total) compared to the existing techniques in MITRE ATT\&CK ($244$ in total).

It is important to mention that we did not carry out a feasibility study for the individual cases, but merely checked whether there was a thematic match.
It is often not possible to evaluate validity one hundred percent accurately, as the subjectivity of the individual evaluators factors in.
The evaluation depends heavily on personal judgment and experience, which can vary greatly from evaluator to evaluator.
Furthermore, different stakeholders may have different priorities and perspectives, leading to different assessments of what constitutes a significant risk or threat.
Also, the context in which risk is assessed, such as organizational culture or industry norms, can influence the perceived severity and likelihood of threats.

\begingroup

\setlength{\tabcolsep}{7pt} 
\begin{figure}[H]
\centering
  \begin{adjustbox}{width=0.4\textwidth}
    \input{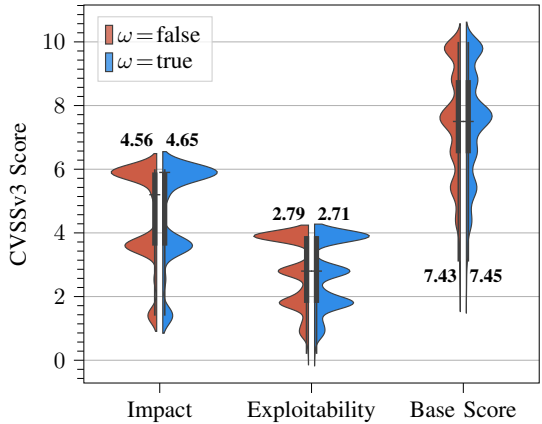}
  \end{adjustbox}
  \caption{Comparison of \gls{cvss} v3 \gls{bsm} with different $\omega$ settings for the T-GEN threat with calculated average scores.}
  \label{fig:test_violin}
\end{figure}
\endgroup

\subsubsection{Effects of Duplicates on Final Scores}
To examine the influence of the $\omega$ parameter, we conducted comparative analyses across various threats.
The results consistently indicated that the uniqueness within the \glspl{cve} minimally affects the overall average score.
In the case of threat T-GEN, we plotted the individual \glspl{cve} scores for impact, exploitability, and base, as shown in Figure~\ref{fig:test_violin}.
The differences between scores from all \glspl{cve} and those considering only unique \glspl{cve} ranged from 0.09 to 0.02.
This suggests that the effect of including duplicate \glspl{cve} in an extraction set is negligible.
Additionally, utilizing $\omega = \text{true}$ substantially reduces the total number of \glspl{cve}—by over 40\% in our example, from 8241 \glspl{cve} to 4928 \glspl{cve}.


\subsection{\gls{cvss} Scores for General O-RAN Threats}
We used the list of generic threats and all threats to illustrate our approach in this paper.
The calculations and the code for the extraction and the rating calculation of all threats can be found in the corresponding Gitlab repository\footnote{https://gitlab.com/submission1660638/orca}.
Table~\ref{tab:o_ran_generic_threats_listing} compares the scores specified by the \gls{oran} Alliance by hand with our extracted and automatically determined, more detailed \gls{cvss} scores. Our approach provides more evidence, as we can determine how we calculate our score. In addition, our approach can be quickly and easily executed programmatically multiple times at different points in time, unlike a manual evaluation.
We compare both the deep and normal methods introduced in Section \ref{sec:scanningmethod}, and denote them as D-\gls{sfc} and N-\gls{sfc}, respectively.
We used the Threat to \glspl{capec} mapping strategy with the \gls{sfc} criterion and calculated the average \gls{cvss} for normal and deep mode with \gls{cvss} v2 and $\omega = false$.
In the following, we provide a brief overview of the limitations of the current methodology and present the initial results of the calculated extraction scores.
However, we do not go into specific threat details, as the focus of this paper is on establishing the framework and tools necessary for conducting such analyses.

\begingroup
\centering
\setlength{\tabcolsep}{6pt} 
\renewcommand{\arraystretch}{1.5} 
\begin{table*}[!ht]
  \caption{Listing of generic threat classes defined by \cite{o_ran_wg11_threat_model} with the added averaged \gls{cvss} v2 information from the extraction using Threat to \glspl{capec}-Mapping with $\cos_{thrs}$=0.55 and soft filter criterion for Normal- and Deep-Mode with $\omega = false$.}
  \label{tab:o_ran_generic_threats_listing}
  \begin{center}
    \begin{tabular}{m{0.105\textwidth}m{0.32\textwidth}|m{0.01\textwidth}m{0.01\textwidth}m{0.02\textwidth}|m{0.02\textwidth}m{0.02\textwidth}m{0.02\textwidth}|m{0.035\textwidth}|m{0.02\textwidth}m{0.02\textwidth}m{0.02\textwidth}|m{0.03\textwidth}}
      \multicolumn{5}{c}{Defined by the O-RAN Alliance \cite{o_ran_wg11_threat_model}} & \multicolumn{4}{c}{Avg. \gls{cvss} for D-SFC}     & \multicolumn{4}{c}{Avg. \gls{cvss} for N-SFC}                                                                                                                                                                                                                                                                                                                               \\
      \cmidrule(lr){1-5}\cmidrule(lr){6-9}\cmidrule(lr){10-13}
      \hline
      {Threat-ID}                                                                      & \multicolumn{1}{c}{{Description}}                 & \vertcell{1.6cm}{Severity}                    & \vertcell{1.6cm}{Likelihood}              & \vertcell{2.6cm}{Risk Score}              & \vertcell{1.6cm}{Impact}      & \vertcell{1.6cm}{Exploitability} & \vertcell{1.6cm}{Base Score}  & {\glspl{cve}} & \vertcell{1.6cm}{Impact}      & \vertcell{1.6cm}{Exploitability} & \vertcell{1.6cm}{Base Score}  & {\glspl{cve}} \\ \hline\hline
      T-AAL                                                                            & Threats concerning Acceleration Abstraction Layer & \color{BrickRed}{\faIcon{circle}}             & \color{YellowOrange}{\faIcon{dot-circle}} & \color{BrickRed}{\faIcon{circle}}         & \cellcolor{YellowOrange}{4.8} & \cellcolor{BrickRed}{7.9}        & \cellcolor{YellowOrange}{5.3} & 3173          & \cellcolor{YellowOrange}{4.7} & \cellcolor{BrickRed}{7.8}        & \cellcolor{YellowOrange}{5.2} & 1900          \\

      T-ADMIN                                                                          & Threats concerning O-Cloud management             & \color{YellowOrange}{\faIcon{dot-circle}}     & \color{BrickRed}{\faIcon{circle}}         & \color{BrickRed}{\faIcon{circle}}         & \cellcolor{YellowOrange}{5.4} & \cellcolor{BrickRed}{7.5}        & \cellcolor{YellowOrange}{5.6} & 12445         & \cellcolor{YellowOrange}{5.3} & \cellcolor{BrickRed}{7.2}        & \cellcolor{YellowOrange}{5.4} & 1580          \\

      T-E2                                                                             & Threats against E2 interface                      & \color{gray}{\faIcon[regular]{circle}}        & \color{gray}{\faIcon[regular]{circle}}    & \color{gray}{\faIcon[regular]{circle}}    & \cellcolor{YellowOrange}{5.5} & \cellcolor{BrickRed}{8.2}        & \cellcolor{YellowOrange}{6.0} & 28824         & \cellcolor{YellowOrange}{5.6} & \cellcolor{BrickRed}{8.4}        & \cellcolor{YellowOrange}{6.2} & 3952          \\

      T-GEN                                                                            & Generic threats against O-Cloud                   & \color{YellowOrange}{\faIcon{dot-circle}}     & \color{BrickRed}{\faIcon{circle}}         & \color{BrickRed}{\faIcon{circle}}         & \cellcolor{YellowOrange}{5.2} & \cellcolor{BrickRed}{7.9}        & \cellcolor{YellowOrange}{5.7} & 81596         & \cellcolor{YellowOrange}{5.2} & \cellcolor{BrickRed}{7.7}        & \cellcolor{YellowOrange}{5.5} & 8038          \\

      T-HW                                                                             & Threats concerning hardware resources             & \color{BrickRed}{\faIcon{circle}}             & \color{BrickRed}{\faIcon{circle}}         & \color{BrickRed}{\faIcon{circle}}         & \cellcolor{YellowOrange}{5.2} & \cellcolor{BrickRed}{7.7}        & \cellcolor{YellowOrange}{5.5} & 276           & \cellcolor{YellowOrange}{5.2} & \cellcolor{BrickRed}{7.7}        & \cellcolor{YellowOrange}{5.5} & 138           \\

      T-IMG                                                                            & Threats concerning VM/Container images            & \color{YellowOrange}{\faIcon{dot-circle}}     & \color{BrickRed}{\faIcon{circle}}         & \color{BrickRed}{\faIcon{circle}}         & \cellcolor{YellowOrange}{5.4} & \cellcolor{BrickRed}{7.7}        & \cellcolor{YellowOrange}{5.7} & 478           & \cellcolor{YellowOrange}{5.5} & \cellcolor{BrickRed}{7.8}        & \cellcolor{YellowOrange}{5.8} & 271           \\

      T-ML                                                                             & Threats against the ML system implementations     & \color{gray}{\faIcon[regular]{circle}}        & \color{gray}{\faIcon[regular]{circle}}    & \color{gray}{\faIcon[regular]{circle}}    & \cellcolor{YellowOrange}{5.4} & \cellcolor{BrickRed}{8.1}        & \cellcolor{YellowOrange}{5.8} & 15886         & \cellcolor{YellowOrange}{5.5} & \cellcolor{BrickRed}{8.0}        & \cellcolor{YellowOrange}{5.9} & 6089          \\

      T-O-CLOUD-ID                                                                     & Threats concerning O-CLoud instance ID            & \color{BrickRed}{\faIcon{circle}}             & \color{BrickRed}{\faIcon{circle}}         & \color{BrickRed}{\faIcon{circle}}         & \cellcolor{YellowOrange}{3.7} & \cellcolor{BrickRed}{7.9}        & \cellcolor{YellowOrange}{4.6} & 6329          & \cellcolor{YellowOrange}{3.7} & \cellcolor{BrickRed}{7.9}        & \cellcolor{YellowOrange}{4.6} & 6329          \\

      T-02                                                                             & Threats concrening O-Cloud interfaces             & \color{Green}{\faIcon{circle-notch}}          & \color{BrickRed}{\faIcon{circle}}         & \color{YellowOrange}{\faIcon{dot-circle}} & \cellcolor{YellowOrange}{5.2} & \cellcolor{BrickRed}{7.7}        & \cellcolor{YellowOrange}{5.5} & 138           & \cellcolor{YellowOrange}{5.2} & \cellcolor{BrickRed}{7.7}        & \cellcolor{YellowOrange}{5.5} & 69            \\

      T-OCAPI                                                                          & Threats concerning the O-Cloud API                & \color{Green}{\faIcon{circle-notch}}          & \color{BrickRed}{\faIcon{circle}}         & \color{YellowOrange}{\faIcon{dot-circle}} & \cellcolor{YellowOrange}{5.2} & \cellcolor{BrickRed}{7.7}        & \cellcolor{YellowOrange}{5.5} & 138           & \cellcolor{YellowOrange}{5.2} & \cellcolor{BrickRed}{7.7}        & \cellcolor{YellowOrange}{5.5} & 69            \\

      T-OPENSRC                                                                        & Threats to open source code                       & \color{BrickRed}{\faIcon{circle}}             & \color{BrickRed}{\faIcon{circle}}         & \color{BrickRed}{\faIcon{circle}}         & \cellcolor{YellowOrange}{4.9} & \cellcolor{BrickRed}{7.9}        & \cellcolor{YellowOrange}{5.4} & 42458         & \cellcolor{YellowOrange}{6.3} & \cellcolor{BrickRed}{8.0}        & \cellcolor{YellowOrange}{6.4} & 11691         \\

      T-PHYS                                                                           & Physical threats                                  & \color{BrickRed}{\faIcon{circle}}             & \color{YellowOrange}{\faIcon{dot-circle}} & \color{BrickRed}{\faIcon{circle}}         & \cellcolor{YellowOrange}{5.4} & \cellcolor{BrickRed}{8.1}        & \cellcolor{YellowOrange}{5.8} & 7996          & \cellcolor{YellowOrange}{6.4} & \cellcolor{BrickRed}{6.9}        & \cellcolor{YellowOrange}{6.0} & 4             \\

      T-RADIO                                                                          & Threats against 5G radio networks                 & \color{BrickRed}{\faIcon{circle}}             & \color{BrickRed}{\faIcon{circle}}         & \color{BrickRed}{\faIcon{circle}}         & \cellcolor{gray!20}{-}        & \cellcolor{gray!20}{-}           & \cellcolor{gray!20}{-}        & 0             & \cellcolor{gray!20}{-}        & \cellcolor{gray!20}{-}           & \cellcolor{gray!20}{-}        & 0             \\

      T-VL                                                                             & Threats concerning the virtualization layer       & \color{BrickRed}{\faIcon{circle}}             & \color{YellowOrange}{\faIcon{dot-circle}} & \color{BrickRed}{\faIcon{circle}}         & \cellcolor{YellowOrange}{4.2} & \cellcolor{BrickRed}{8.3}        & \cellcolor{YellowOrange}{5.1} & 3644          & \cellcolor{YellowOrange}{4.5} & \cellcolor{BrickRed}{7.8}        & \cellcolor{YellowOrange}{5.1} & 724           \\

      T-VM-C                                                                           & Threats concerning VMs/Containers                 & \color{BrickRed}{\faIcon{circle}}             & \color{BrickRed}{\faIcon{circle}}         & \color{BrickRed}{\faIcon{circle}}         & \cellcolor{YellowOrange}{5.4} & \cellcolor{BrickRed}{7.6}        & \cellcolor{YellowOrange}{5.6} & 19632         & \cellcolor{YellowOrange}{5.8} & \cellcolor{BrickRed}{6.7}        & \cellcolor{YellowOrange}{5.5} & 2685          \\

      T-Y1                                                                             & Threats against Y1 interface                      & \color{gray}{\faIcon[regular]{circle}}        & \color{gray}{\faIcon[regular]{circle}}    & \color{gray}{\faIcon[regular]{circle}}    & \cellcolor{YellowOrange}{4.9} & \cellcolor{BrickRed}{8.1}        & \cellcolor{YellowOrange}{5.5} & 16153         & \cellcolor{YellowOrange}{4.7} & \cellcolor{BrickRed}{8.1}        & \cellcolor{YellowOrange}{5.3} & 13728         \\  \hline
    \end{tabular}

  \end{center}
  \begin{center}
    \vspace{1em}
    O-RAN Scores: \color{BrickRed}{\faIcon{circle}} $\equiv$ High, \color{YellowOrange}{\faIcon{dot-circle}} $\equiv$ Medium, \color{Green}{\faIcon{circle-notch}} $\equiv$ Low, \color{gray}{\faIcon[regular]{circle}} $\equiv$ None \color{Black}| Avgerage \gls{cvss}: \color{BrickRed}{$10 \dotsm 6.67 $} $\equiv$ High, \color{YellowOrange}{$6.\overline{6} \dotsm 3.34 $} $\equiv$ Medium, \color{Green}{$3.\overline{3} \dotsm 0$} $\equiv$ Low
  \end{center}
\end{table*}
\endgroup

\begin{figure*}[tp]

  \centering

  \minipage{0.48\textwidth}

  \begin{adjustbox}{width=88mm}
    \includegraphics[width=.9\textwidth]{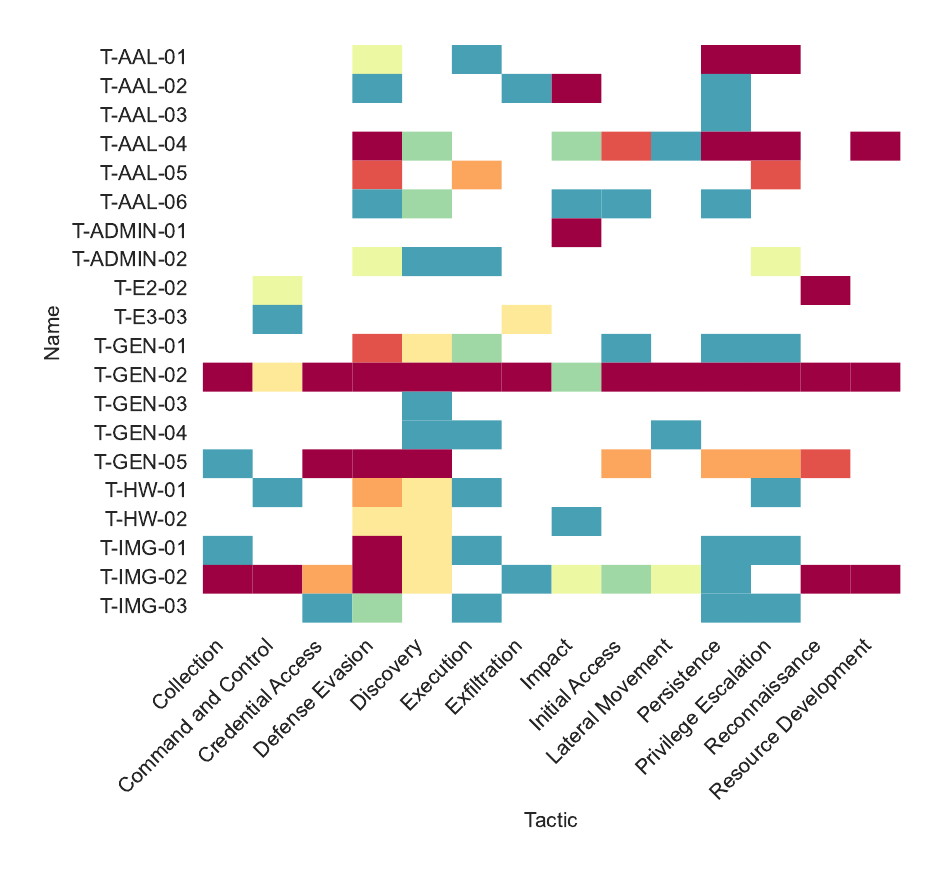}
  \end{adjustbox}

  \endminipage\hfill
  \minipage{0.52\textwidth}

  \begin{adjustbox}{width=96mm}
    \includegraphics[width=.9\textwidth]{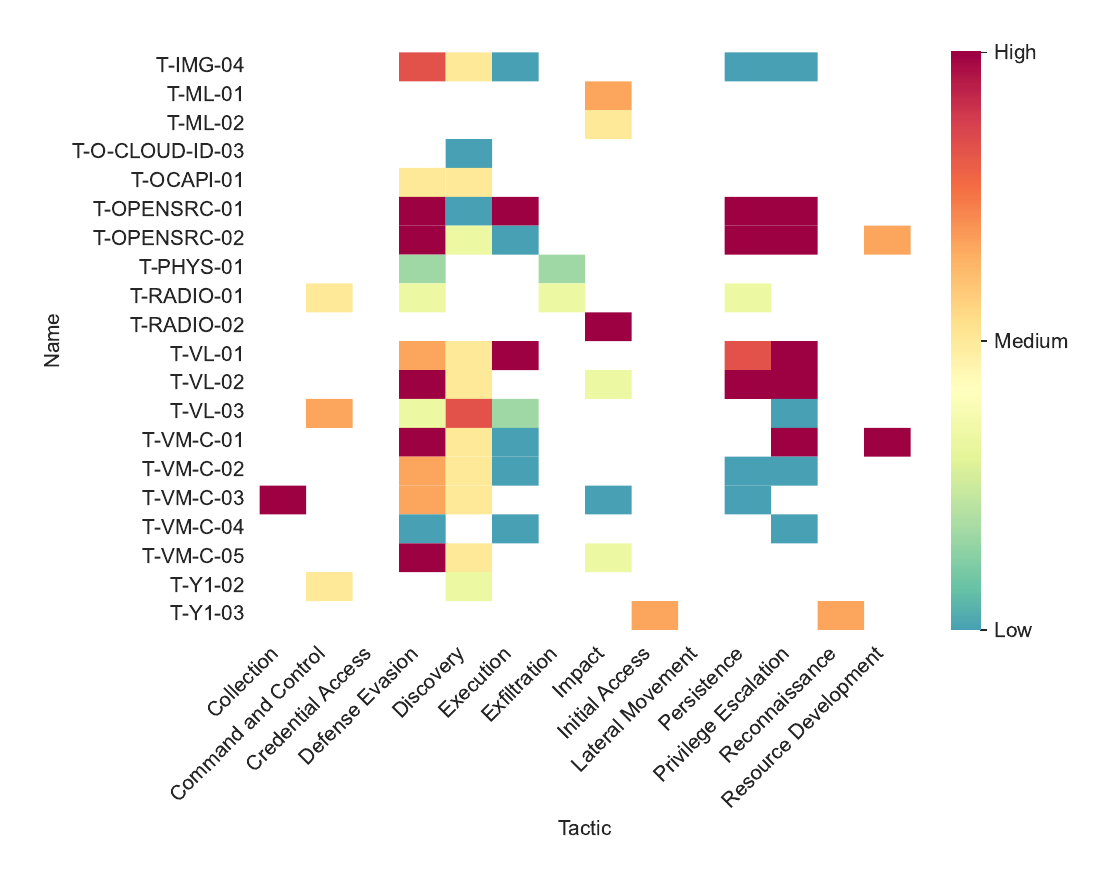}
  \end{adjustbox}

  \endminipage\hfill

  \caption{Heat map of accumulated identical tactics per generic threat group classification for \gls{cvss} v2 information from the extraction using Threat to Techniques-Mapping with $\cos_{thrs}$=0.50 and soft filter criterion for Normal-Mode with $\omega = true$.}
  \label{fig:tactics_headmap}

\end{figure*}

\subsubsection{Prevailing method}
The risk assessment process described by the O-RAN Alliance's Security \gls{wg} 11 in \cite{o_ran_wg11_threat_model} comprises several key steps.
These include the identification of assets, threats, and vulnerabilities, followed by an assessment of the associated risk.
The assessment is categorized by the criticality of the threat and the likelihood of its occurrence.
A score is then determined using the RISK formula ($Severity \times Likelihood$).
The individual assessment levels are divided into three categories: low, medium, and high, with low representing the lowest and high the highest degree of severity.
However, it should be noted that this approach to determining risk is not without limitations.
This type of traditional risk assessment is highly susceptible to the subjectivity of the individual being assessed.
Due to the manual assessment process, it is prone to varying interpretations, introducing a great degree of subjectivity.
To solve these difficulties, our approach uses a systematic, iterative, repeatable risk ranking based on currently existing vulnerabilities, which are assessed through empirical observations of adversary behavior.

\subsubsection{First Pre-analysis of the Results}
The first thing that stands out when comparing only the two extraction modes in one \gls{cvss} version of our approach is that the average \gls{cvss} classes do not differ much.
There are only three differences in the High and Low classes when looking at the different \gls{cvss} results. These slight variations are due to the fact that the D-\gls{sfc} additionally extracts the related \glspl{capec} and thus provides more comprehensive analysis possibilities.
This fact can also be determined by comparing the total number of mapped \glspl{cve}.
In some cases, the difference between N-\gls{sfc} and D-\gls{sfc} ranges by a multitude.
Due to the larger extraction area, some of the \glspl{cve} in the endset are not as heavy, which results in a weakening of the average scoring. However, an opposite movement is also recognizable, in that more severe \glspl{cve} lead to an increase. Generally speaking, the more detailed and comprehensive you want the evaluation to be, the more you decide for or against the deep mode.

It is noteworthy that the evaluation of exploitability differs significantly between \gls{cvss} versions 2 and 3, with scoring classes shifting from high in v2 to low in v3.
This shift does not indicate a sudden decrease in the exploitability or relevance of individual \glspl{cve} but rather reflects changes in the scoring methodology used by \gls{cvss}. As previously discussed, selecting the appropriate \gls{cvss} version is critical for accurate analysis and is therefore of fundamental importance. However, this decision must be made by the implementing entity based on key factors specific to the use case.

\subsubsection{Summary of our Scoring}
Overall, our approach demonstrates a higher degree of granularity and presents a robust methodology for the detailed and iterative analysis of individual threat classes, as we further elaborate in Section \ref{sub:further_threat_analysis}.
This enables users to obtain a rapid, comprehensive, and detailed overview of specific threats, thereby facilitating informed decision-making and subsequent action.

\subsection{Further Threat Analysis Possibilities}
\label{sub:further_threat_analysis}
The extracted datasets provide a broad array of opportunities for further analysis, which are closely aligned with the specific requirements of the implementing entities.
In general, we categorize further analysis techniques into two types: \textit{specific detailed methods}, such as averaged spider plots of a single threat, time course of \glspl{cve} for a threat; and \textit{broader comprehensive approaches}, such as accumulated base score ratings divided by severity level, and heat map tactics coverage.
To illustrate, we present an example implementation of tactics coverage across various threats, chosen for its visual simplicity and clarity, making it a straightforward and self-explanatory example.

Our extraction process enables not only the calculation of threat scores but also a wide range of additional analyses.
For instance, the coverage of individual \gls{attack} tactics per threat can be visually represented.
As an example, we have applied this to the threats of the generic threat group, with the results depicted in Figure~\ref{fig:tactics_headmap}.
This visualization allows for easy identification of potentially dangerous threats.
Notably, the threat class T-GEN-02 stands out due to its high cumulative score across nearly all tactics, indicating a higher likelihood of success for attackers across various attack tactics.
Overall, this approach serves as a valuable tool for prioritizing mitigation efforts and more effectively identifying and localizing specific problem areas.

\section{Conclusion}
\label{sec:conclusion}

Vulnerability assessment is a complex discipline that must address numerous hypothetical threats and events. The analysis is frequently subject to significant influence from subjective evaluations, as state-of-the-art methods primarily rely on manual analysis. To mitigate this challenge, in this paper, we employed a quantitative, automated, and iteratively repeatable methodology. We presented the methodology and applied it in the context of \gls{oran} software-based networks. While some degree of uncertainty may persist in existing approaches, the insights derived from our method provide quantitative assessments of the threat severity and provide efficient numerical evaluations in complex \gls{oran} scenarios. This framework enables a more precise evaluation of current components, informed planning and prioritization of future initiatives, and a more accurate assessment of overall economic risk.


Future work in this area to enhance our current methodology may include the automated generation and execution of test cases for specific groups of techniques. One possible extension could involve integrating our approach with the efforts of the Atomic Red Team \cite{atomicredteam_coverage}, which has developed a library of tests mapped to the \gls{attack} framework. This integration could facilitate the efficient and reproducible assessment of environments for technique presence through standardized testing. This integration would be a major step forward in the automated security testing of system environments and deployments. By enabling automated active testing, it would enhance the speed and reliability with which the security of specific components can be verified.

\newpage
\bibliographystyle{IEEEtran}
\bibliography{biblio}

\clearpage

\twocolumn[{
\begin{center}
  \label{tab:extraction_table}
  \captionof{table}{Example from the extracted data frame for threat T-GEN-02 for CAPEC-122}
  \resizebox{\textwidth}{!}{
    \begin{tabular}{|l|l|l|l|l|l|l|l|}
      \hline
      threat\_id & cve\_id          & cwe\_id & capec\_id & v2\_impactScore & v2\_exploitabilityScore & v2\_baseScore & published           \\ \hline
      T-GEN-02   & CVE-2017-16757   & CWE-732 & CAPEC-122 & 6.4             & 3.9                     & 4.6           & 2017-11-09 21:29:00 \\ \hline
      T-GEN-02   & CVE-2017-1000403 & CWE-732 & CAPEC-122 & 6.4             & 8                       & 6.5           & 2018-01-26 02:29:01 \\ \hline
      ...        & ...              & ...     & ...       & ...             & ...                     & ...           & ...                 \\ \hline
      T-GEN-02   & CVE-2022-33710   & CWE-269 & CAPEC-122 & 10              & 3.9                     & 7.2           & 2022-07-12 14:15:18 \\ \hline
      T-GEN-02   & CVE-2024-23620   & CWE-269 & CAPEC-122 & 6.4             & 10                      & 7.5           & 2023-11-13 16:15:28 \\ \hline
    \end{tabular}
  }
\end{center}
}]

\appendix
\section{Appendix}

\subsection{Example for the individual Pipeline Steps}
\label{app:sec:example}
To gain a more thorough understanding of our methodology, in which each step is performed to progress from the initial threat data to the final evaluations, we present a simplified example utilizing the threat \textit{T-GEN-02}. This example is based on the \gls{ttcm} branch with N-\gls{hfc}, $\omega = true$ and a $cos_{thrs}$ of 0.55.

\subsubsection{Threat Description - Input}
The initial phase typically involves a substantial amount of data related to the specific threat. In our example, this information is represented in Listing \ref{lst:threat_description} using the \gls{json} format.
\begin{lstlisting}[caption={\gls{json} content of the threat T-GEN-02},language=json,firstnumber=1,label={lst:threat_description}]
{
"Threat ID":"T-GEN-02",
"Threat title":"Malicious access to exposed services using valid accounts",
"Threat Description":"Access to valid accounts to use the O-Cloud services is often a requirement, which could be obtained through credential pharming or by obtaining the credentials from users after compromising the network. ...  Access may be also gained through an exposed service that doesn't require authentication. In containerized environments, this may include an exposed Docker API, Kubernetes API server, kubelet, or web application such as the Kubernetes dashboard.",
"Threat agent":"All",
"Vulnerability":[
    "Lack of authentication"
],
"Threatened Asset":"ASSET-D-12, ASSET-D-13, ASSET-D-14, ASSET-D-15, ASSET-D-16, ASSET-D-17, ASSET-D-18, ASSET-D-19, ASSET-D-20, ASSET-D-29, ASSET-D-31, ASSET-D-32",
"Affected Components":"O-Cloud, Apps/VNFs/CNFs"
}
\end{lstlisting}

\subsubsection{Threat Summary - Preprocessing}
The preprocessing step is performed prior to the execution of the two potential branches. In our example, a text paragraph (Listing \ref{lst:preprocessing}) is generated and subsequently utilized as input for the \gls{nlp} model.

\begin{lstlisting}[caption={Preprocessed text paragraph},language=json,firstnumber=1,label={lst:preprocessing}]
"summary": A Threat with the title Malicious access to exposed services using valid accounts and the description Access to valid accounts to use the O-Cloud services is often a requirement, which ... this may include an exposed Docker API, Kubernetes API server, kubelet, or web application such as the Kubernetes dashboard.
\end{lstlisting}

\subsubsection{Threat Summary - Embeddings}
Listing \ref{lst:embeddings} displays the embeddings produced by the sentence transformer. These embeddings correspond to a 384-dimensional dense vector, which is subsequently utilized to compute the cosine similarity.
\begin{lstlisting}[caption={Generated embeddings by the language model},language=json,firstnumber=1,label={lst:embeddings}]
"embeddings": [-2.05694959e-02 -4.06922102e-02 -1.01613447e-01 -7.79516669e-03
  4.01788913e-02 -3.49763446e-02  6.36101142e-02  4.13039252e-02
    ...
  2.37159394e-02  4.14265916e-02  5.60964905e-02 -2.71430593e-02]
\end{lstlisting}

\subsubsection{Threat - Mapping}
The cosine similarity is subsequently computed between all \glspl{capec} embeddings and the generated embeddings of the threat shown in Listing \ref{lst:embeddings}. Based on the parameters used in this example, six \glspl{capec} are identified to have a higher similarity then 0.55. The respective \gls{capec} ID is output alongside the threat title, domain, and the calculated cosine similarity, as demonstrated in Listing \ref{lst:calculated_mappings}.
\begin{lstlisting}[caption={Calculated mappings for the given threat},language=json,firstnumber=1,label={lst:calculated_mappings}]
T-GEN-02;enterprise-attack;Malicious access to exposed services using valid accounts;CAPEC-122;0.5660987
T-GEN-02;enterprise-attack;Malicious access to exposed services using valid accounts;CAPEC-114;0.5621336
T-GEN-02;enterprise-attack;Malicious access to exposed services using valid accounts;CAPEC-560;0.5553874
T-GEN-02;enterprise-attack;Malicious access to exposed services using valid accounts;CAPEC-555;0.5601635
T-GEN-02;enterprise-attack;Malicious access to exposed services using valid accounts;CAPEC-510;0.6071962
T-GEN-02;enterprise-attack;Malicious access to exposed services using valid accounts;CAPEC-600;0.5650064
\end{lstlisting}

The \glspl{capec} found are:
\begin{itemize}
  \item CAPEC-122: Privilege Abuse
  \item CAPEC-114: Authentication Abuse
  \item CAPEC-560: Use of Known Domain Credentials
  \item CAPEC-555: Remote Services with Stolen Credentials
  \item CAPEC-510: SaaS User Request Forgery
  \item CAPEC-600: Credential Stuffing
\end{itemize}
It is immediately noticeable that the majority of the identified results closely align with the given threat description and title.

\subsubsection{Extraction}
The associated \glspl{cwe} and subsequently the \glspl{cve} must be extracted from all identified \glspl{capec}. Considering all \glspl{cve} generated since their inception often results in a substantial number of individual \glspl{cve}. Table \ref{tab:extraction_table} provides an excerpt from the pickle file, which is saved after extraction. For instance, \gls{capec}-122 includes a total of 1,823 distinct \glspl{cve}. By calculating the average of all scores, we derive an impact score of 6.39, an exploitability score of 6.17, and a base score of 5.66 for the threat \textit{T-GEN-02}.

\end{document}